\documentclass[]{revtex4-2}
\usepackage[utf8]{inputenc}

\usepackage{graphicx}
\usepackage{amsfonts}
\usepackage{amsmath}
\usepackage{amssymb}
\usepackage{epstopdf}
\usepackage{color}
\usepackage{bm} 
\usepackage{appendix}
\begin{document}

\title{Modulational electrostatic wave-wave interactions in plasma fluids 
        modeled by asymmetric coupled nonlinear Schr\"odinger (CNLS) equations
}
\author{
N. Lazarides$^1$, 
Giorgos P. Veldes$^2$, 
Amaria Javed$^3$,
Ioannis Kourakis$^{1, 4, 5}$ \\ \ \\ 
$^1$ Department of Mathematics, Khalifa University of Science and Technology, \\
     P.O. Box 127788, Abu Dhabi, UAE \\
$^2$ Department of Physics, University of Thessaly, Lamia 35100, Greece \\
$^3$ Center for Quantum and Topological Systems, \\
      New York University Abu Dhabi, 
      P.O. Box 129188, Abu Dhabi, UAE \\
$^4$ Space and Planetary Science Center, 
    Khalifa University of Science and Technology, \\
    P.O. Box 127788, Abu Dhabi, UAE
    \\
$^5$ National and Kapodistrian University of Athens, Faculty of Physics, \\ 
     School of Science, Panepistimiopolis, Zografos 157 84 Athens, Greece 
}
\date{\today}
%*****************************************************************************80
\begin{abstract}
The interaction between two co-propagating electrostatic wavepackets 
characterized by arbitrary carrier wavenumber is considered. A one-dimensional 
(1D) non-magnetized plasma model is adopted, consisting of a cold inertial ion 
fluid evolving against a thermalized (Maxwell-Boltzmann distributed) electron 
background. A multiple-scale perturbation method is employed to reduce the 
original model equations to a pair of coupled nonlinear Schr\"odinger (CNLS) 
equations governing the dynamics of the wavepacket amplitudes (envelopes). 
The CNLS equations are in general asymmetric for arbitrary carrier wabvenumbers. 
Similar CNLS systems have been derived in the past in various physical contexts, 
and were found to support soliton, breather, and rogue wave solutions, among 
others. A detailed stability analysis reveals that modulational instability (MI) 
is possible in a wide range of values in the parameter space. The instability 
window and the corresponding growth rate are determined, considering different 
case studies, and their dependence on the carrier and the perturbation wavenumber 
is investigated from first principles. Wave-wave coupling is shown to favor MI 
occurrence by extending its range of occurrence and by enhancing its growth rate. 
Our findings generalize previously known results usually associated with 
symmetric NLS equations in nonlinear optics, though taking into account the 
difference between the different envelope wavenumbers and thus group velocities. 
\end{abstract}
%*****************************************************************************80
\maketitle

%*****************************************************************************80
\section{Introduction}
%*****************************************************************************80
The theoretical and experimental investigation of plasma dynamics has grown 
significantly in the last decades, because of their vast application prospects 
related  to Space and fusion plasmas, among others. The intrinsic nonlinear 
properties of plasmas as dispersive media are known to play a crucial role, 
leading to various types of collective excitations in the form of localized modes 
(solitary waves, shocks). Plasma fluid models, mimicking Navier-Stokes equation 
in hydrodynamics, can be reduced through standard methods to well-known nonlinear 
partial differential equations of mathematical physics, such as the celebrated 
nonlinear Schr{\"o}dinger (NLS) equation known to govern the envelope dynamics of 
a modulated wavepacket in dispersive media.

Following the pioneering work of Refs. \cite{Shimizu1972,Kakutani1974}, the NLS 
equation has subsequently been derived for a variety of electrostatic plasma 
modes within the fluid plasma formalism. Building up on similar considerations in 
nonlinear optics, its localized solutions (envelope solitons, breathers) have in 
fact been associated with freak waves 
\cite{Kourakis2005,Chowdhury2018, Singh2020, Sarkar2020}. 
Similar derivations of formally analogous NLS equations (but in different context) 
have focused on electromagnetic waves in plasmas, modeled by fluid-Maxwell 
equations \cite{Borhanian2009,Borhanian2015,Veldes2013b}, 
also proposing  a rogue wave related interpretation \cite{Veldes2013b}.

If one considers a pair of co-propagating (interacting) wavepackets in plasmas, 
a particular plasma fluid model leads to a pair of coupled NLS  (CNLS) equations, 
whose coefficients generally depend on the carrier wavenumber(s) of the 
respective waves. Examples include electron plasma (Langmuir) waves and ion 
acoustic waves mainly 
\cite{Spatschek1978,Som1979,Kofane2020}, 
amidst other studies that focused mostly on electromagnetic modes  
\cite{McKinstrie1989,McKinstrie1990,Luther1990,Luther1992,Kourakis2005b,Singh2013,Borhanian2017}.

%*****************************************************************************80
A number of works on single NLS equations derived from various plasma fluid 
models have appeared in the literature, in which soliton, breather, 
and rogue wave solutions were investigated by using known analytical forms 
representing those localized structures. However, analogous investigations of 
two (or more) nonlinear co-propagating and interacting waves in plasma systems 
are rather scarce (see the references above). Thus, it is meaningful to derive 
a pair of coupled NLS equations for the plasma fluid model described above, 
showcasing the analytical dependence of all relevant coefficients on the carrier 
wavenumber(s) of the respective waves. It is also meaningful and, in fact, a 
challenging task to investigate how the modulational instability phenomenon, a 
universal mechanism responsible for wave localization, may occur such 
electrostatic wave pairs.

Although the derived CNLS equations obey a general form, the dependence of the  
dispersion, self-nonlinearity and cross-modulation coefficients on the wavenumbers 
of the respective waves (and, in fact, on the plasma parameters) is unique for 
each considered plasma fluid model. An important common feature in is that, 
assuming arbitrary carrier wavenumber (values), these coefficients do not exhibit 
any obvious symmetry, thus rendering the CNLS equations asymmetric. These general 
CNLS equations may formally reduce, in certain special case, to known CNLS models 
such as the Manakov model \cite{Manakov1974,Radhakrishnan1995} which is long known 
to be integrable. 

Various plasma related experimental studies have actually focused on the emergence 
of envelope solitons \cite{Bailung2011} and breathers \cite{Pathak2016}. It is 
remarkable that the observed soliton and breather waveforms can be investigated 
within the framework of the NLS equation and its already known analytical solutions. 
Recall, at this point, that CNLS equations and their known solutions have been 
employed in the description of solitons and breathers in water waves, exhibiting 
good agreement with experimental observations \cite{He2022}. In the light of the 
above considerations, a meticulous study of the modulational instability 
mechanism associated with energy localization in plasmas may be worth 
investigating in laboratory realizations of coupled electrostatic waveforms, 
by means of CNLS (systems of) equation whose coefficients encompass all of the 
intrinsic plasma wave dynamics. 
%*****************************************************************************80

In this paper, a pair of CNLS equations is derived from a plasma model 
comprising of a cold inertial ion fluid evolving against a thermalized electron 
background that follows the Maxwell-Boltzmann distribution, using a standard multiple 
scale approach (the Newell method). The coefficients of the CNLS equations are 
provided as functions of the wavenumbers of the two interacting waves and the 
coefficients of the (approximation of) the Maxwell-Boltzmann distribution. The 
values of these coefficients for arbitrary carrier wavenumbers of the two waves  
render the system generally asymmetric. A compatibility condition is also 
derived, in the form of a fourth degree polynomial in the frequency of the 
perturbation. The modulational instability (MI) in the system is then investigated 
with respect to the wavenumber (values) of the two carrier waves and the 
wavenumber of the perturbation, by numerically obtaining the roots of the 
compatibility condition and by calculating the magnitude of the associated 
instability growth rate.

This Article is laid out as follows: the following Section 2 introduces the plasma 
fluid model equations, from which the pair of CNLS equations are derived using the 
multiple-scale perturbation method. In Section 3, we undertake an analytical 
study of the modulational stability analysis of plane wave solutions of the CNLS 
equations. A detailed parametric analysis of modulational instability occurrence 
is carried out, based on a numerical calculation of the growth rate numerically, 
in Section 4. Our main results are summarized in the concluding Section 5.

%*****************************************************************************80
\section{Plasma fluid model and reduction to CNLS equations}
%*****************************************************************************80
\subsection{Model equations and normalization}

A one-dimensional (1D), non-magnetized plasma model is considered for 
electrostatic excitations, consisting of a cold inertial ion fluid of density 
$n_i =n$ and velocity $u_i =u$, evolving against a thermalized electron 
background. The electron component follows the Maxwell-Boltzmann distribution,
hence its (normalized) density $n_e$ is given by 
\begin{equation}
\label{eq00}
   n_e  = e^{\phi} \simeq 1 +c_1 \phi +c_2 \phi^2 +c_3 \phi^3 ,
\end{equation}
where a Mc Laurin series expansion in powers of the electrostatic potential 
$\phi$ was considered in the last step, for $|\phi| \ll 1$. The constant 
coefficients for the Maxwell-Boltzmann distribution are $c_1 =2 c_2 =6 c_3 =1$.

That plasma fluid model comprises the following continuity, momentum and Poisson 
equation(s)
\begin{eqnarray}
\label{eq01}
   \frac{\partial n}{\partial t} +\frac{\partial (n u)}{\partial x} =0, 
   \qquad
%\label{eq02}
   \frac{\partial u}{\partial t} +u \frac{\partial u}{\partial x} =
   -\frac{\partial \phi}{\partial x}, 
   \qquad
%\label{eq03}
   \frac{\partial^2 \phi}{\partial x^2} \simeq 1 -n +c_1 \phi +c_2 \phi^2 +c_3 \phi^3,
\end{eqnarray}
respectively, where $n$, $u$, and $\phi$ are functions of space $x$ and time $t$.

The above equations (\ref{eq00}) and (\ref{eq01}) are given in normalized units. 
Specifically, the space and time variables $x$ and $t$ are normalized to the 
inverse ion plasma frequency $\omega_{pi}^{-1}$ and to the Debye screening length 
$\lambda_D$, respectively, with 
\begin{equation}
\label{eq03.2}
   \lambda_D =\left(\frac{\varepsilon_0 k_B T_e}{z_i n_{i0} e^2}\right)^{1/2}, 
   \qquad {\rm and} \qquad 
   \omega_{pi}=\left[\frac{n_{0, i} (z_i e)^2}{\varepsilon_0 m_i}\right]^{1/2} \, .
\end{equation}

The number density variable(s) of ions $n$ and electrons $n_e$ are normalized 
to their respective equilibrium values $n_{i, 0}$ and $n_{e, 0} = z_i n_{i, 0}$, 
the ion velocity $u$ is normalized to 
$c_s = \lambda_D \, \omega_{pi} =\left(\frac{z_i k_B T_e}{m_e}\right)^{1/2}$, i.e.  
essentially the (ion-acoustic) sound speed. Finally, the electrostatic potential 
is normalized to $k_B T_e/e$. 
Adopting a standard  notation, the symbols $\varepsilon_0$, $k_B$, $T_e$, $e$, 
$m_i$, $z_i$ appearing in the latter expressions denote respectively the 
dielectric permittivity in vacuum, the Boltzmann's constant, the (absolute) 
electron temperature, the electron charge, the ion mass, and the degree of 
ionization ($z_i = q_i/e$, where $q_i$ is the ion charge). 
%*****************************************************************************80

%*****************************************************************************80
\subsection{Derivation of the coupled CNLS equations}

We have undertaken a long algebraic procedure, adopting a multiple scales 
perturbation technique, to derive a system of evolution equations for the 
amplitudes (envelopes) describing a pair of (co-propagating) electrostatic 
wavepackets. Details on the methodology adopted are reported in the Appendix. 
For clarity in presentation, only the main steps are presented in the following.

We introduce fast and slow variables, viz. $x_n =\varepsilon^n x$ and 
$t_n =\varepsilon^n t$ (for $n = 0,1,2,3,...$), where $\varepsilon \ll 1$ is a 
small real constant. On the other hand,  
the dependent variables $n$, $u$, and $\phi$ are expanded in powers of 
$\varepsilon$, around the equilibrium state, as
\begin{equation}
\label{eq07new}
   n =1 +\varepsilon n_1 +\varepsilon^2 n_2 +\varepsilon^3 n_3 +\cdots, \qquad
   u =\varepsilon u_1 +\varepsilon^2 u_2 +\varepsilon^3 u_3 +\cdots, \qquad 
   \phi =\varepsilon \phi_1 +\varepsilon^2 \phi_2 +\varepsilon^3 \phi_3 +\cdots.
\end{equation}
Substituting the above expansions into Eqs. (\ref{eq01}), one obtains various 
sets of equations at different orders in $\varepsilon$. We have proceeded up to 
order $\varepsilon^3$.

The first-order ($\sim \epsilon^1$) equations are solved by  
introducing  the following {\em Ansatz}:
 \begin{eqnarray}
\label{eq_i04new}
  S_1 =S_{1,1}^{(-1)} e^{-i \theta_1} +S_{1,1}^{(1)} e^{+i \theta_1} 
         +S_{1,2}^{(-1)} e^{-i \theta_2} +S_{1,2}^{(1)} e^{+i \theta_2},
\end{eqnarray}
where $S_1 = \{n_1, u_1, \phi_1\}$ is the state vector (triplet) to 1st order, 
and the phases are given by $\theta_j =k_j x -\omega_j t$, with $k_j$ and 
$\omega_j$ being the wavevector and the corresponding frequency of the $j-$th 
wave; note that reality of the state variables imposes 
$S_{1,j}^{(-1)} = (S_{1,j}^{(1)})^*$ (for $j=1,2$), where the star (*) denotes 
the complex conjugate. 
A compatibility condition leads to a (common) linear frequency dispersion 
relation, to be obeyed by both wavepackets ($j=1,2$). The angular frequency 
for the $j-$th wave and  the corresponding group velocity read: 
\begin{equation}
\label{eq_i17new}
   \omega_j =\frac{k_j}{\sqrt{k_j^2 +c_1}}, 
   \qquad 
   v_{g,j} =\frac{\partial \omega_j}{\partial k_j} =\frac{c_1}{(k_j^2 +c_1)^{3/2}}.
\end{equation}
(Note that only the positive branch for both $\omega$ and $k$ is considered.)
In the following, both the wavenumbers $k_j$ and the corresponding dispersion 
frequency $\omega_j$ will appear in the derived mathematical expressions. This 
is done only for convenience and for ease in the presentation and wherever it 
occurs, it is implied that $\omega_j$ is provided by Eq. (\ref{eq_i17new}) as 
a function of $k_j$.

The first-order equations may now be solved in terms of the variables 
$\phi_{1,j}^{(1)} =\Psi_j$ (for $j=1,2$), to yield 
\begin{equation}
\label{eq_i17.2new}
    n_{1,j}^{(1)} =\left(\frac{k_j}{\omega_j}\right)^2 \Psi_j, 
    \qquad
    u_{1,j}^{(1)} =\frac{k_j}{\omega_j} \Psi_j,
    \qquad 
    \phi_{1,j}^{(1)} =\Psi_j \, ,
\end{equation}
where we defined the function(s) $\Psi_j$ (for the $j-$ wave), denoting the 
leading-order electrostatic amplitude disturbance (amplitudes). These will be 
our main state variables in the modulational analysis to follow. 

%****************************************************************************fig
% Figure 1...
\begin{figure}[htp]
    \centering
    \includegraphics[width=12cm]{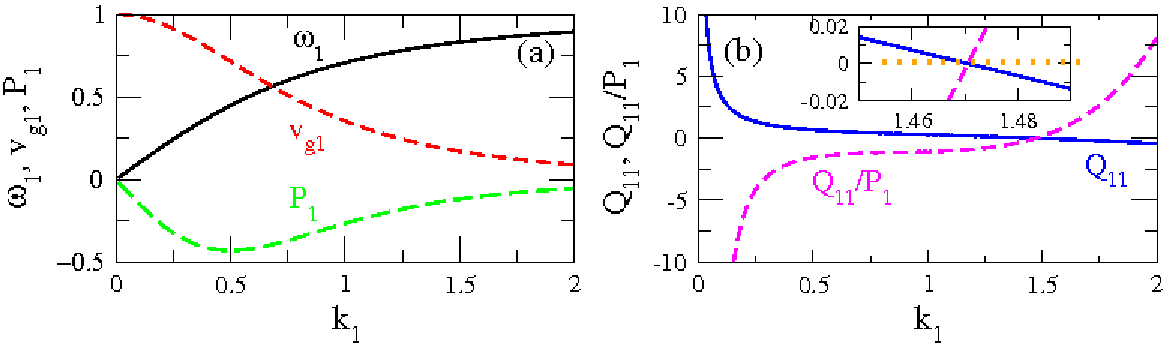}
    \caption{
    (a) The linear frequency dispersion $\omega_1$, the group velocity $v_{g,1}$, 
    and the dispersion coefficient $P_1$ are depicted as functions of the 
    wavenumber $k_1$.
    Note that $P_1$ is always negative (in the cold fluid model).
    (b) The nonlinearity coefficient $Q_{11}$ and the ratio $Q_{11}/P_1$ are 
    depicted as functions of the wavenumber $k_1$. Note that both $Q_{11}$ and 
    $Q_{11}/P_1$ can be either positive or negative, and that they have a root 
    at $k_1 =1.47$ as was expected. 
    }
    \label{fig1}
\end{figure}
%****************************************************************************fig

An interesting constraint arises in  second order ($\sim \varepsilon^2$), where 
suppression of secular terms imposes the condition(s) 
\begin{equation}
\label{eq15new}
   \frac{\partial \Psi_j}{\partial t_1} =-v_{g,j} \frac{\partial \Psi_j}{\partial x_1} \, 
\end{equation} 
(for $j = 1, 2$).  
Recall that the group velocity was defined  earlier, based on the dispersion 
relation found earlier. Physically speaking, this means that the wavepacket 
amplitudes move at the group velocity, as expected. 
This is a well known result, related with the so called Newell technique in 
nonlinear optics; see e.g. in Ref. \cite{Infeld-Rowlands}.
%*****************************************************************************80

%****************************************************************************fig
% Figure 2...
\begin{figure}[htp]
    \centering
    \includegraphics[width=7.5cm]{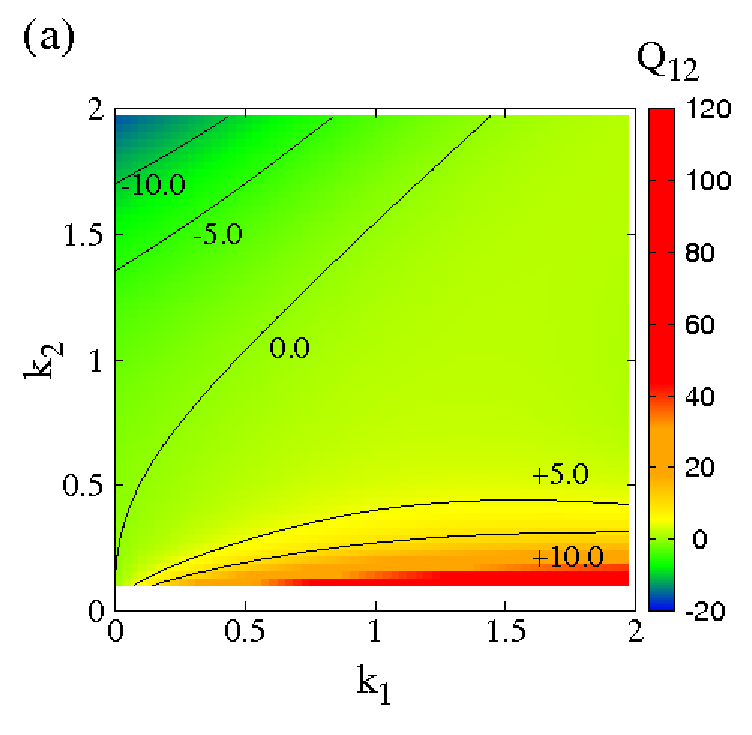}
    \includegraphics[width=7.5cm]{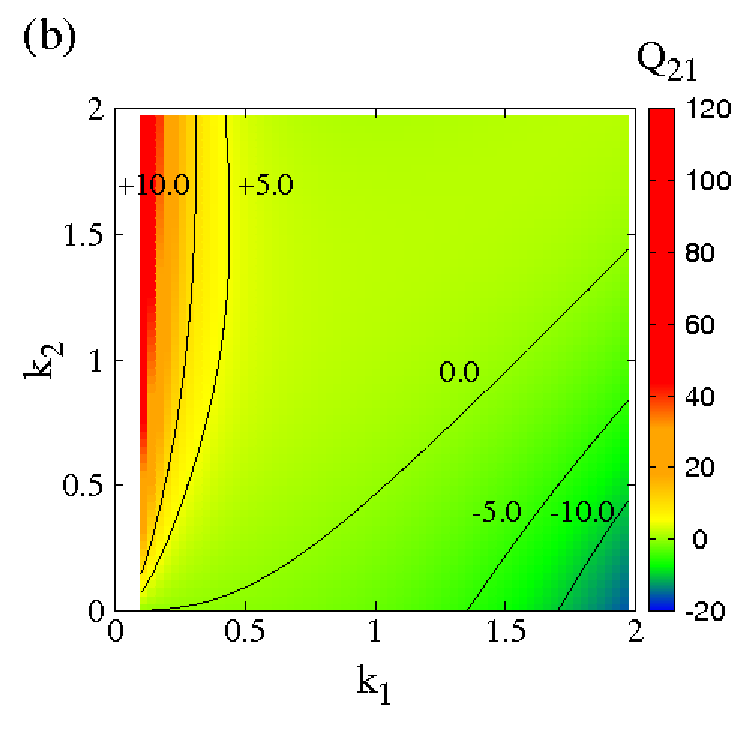} \\
    \includegraphics[width=7.5cm]{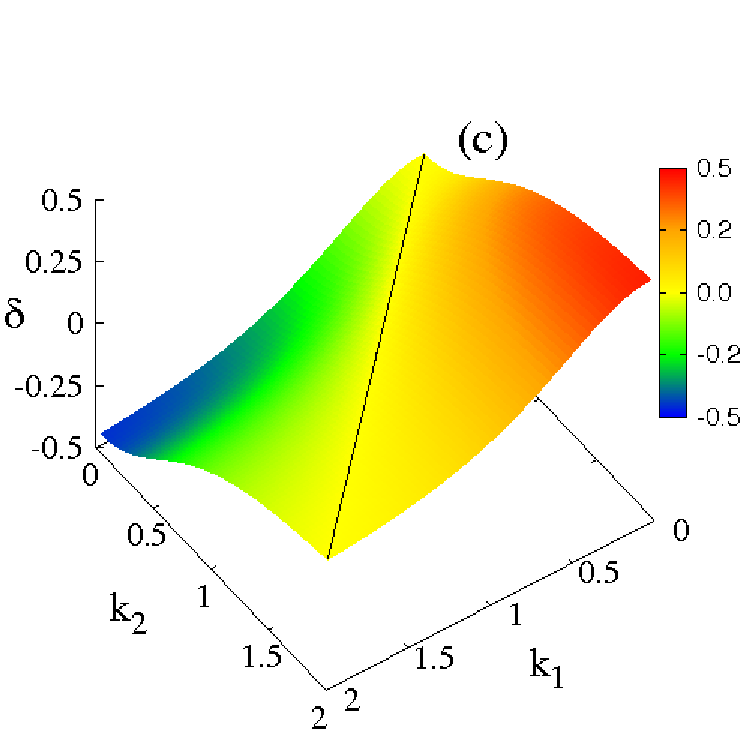}
    \includegraphics[width=7.5cm]{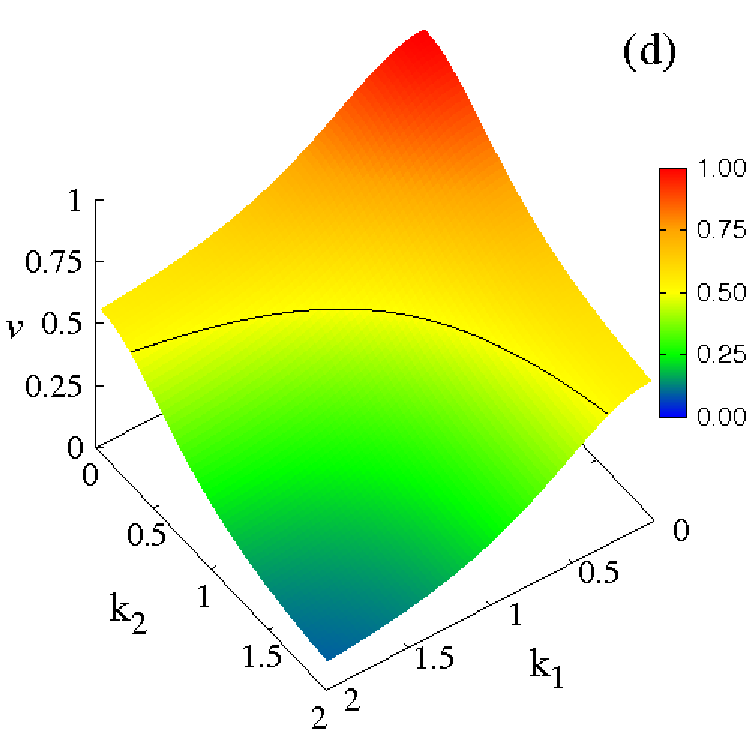}
    \caption{
    The nonlinear coupling coefficient $Q_{12}$ (a) and $Q_{21}$ (b) are 
    depicted as functions of the respective wavenumber(s) of the two 
    co-propagating carrier waves, $k_1$ and $k_2$. The black curves in (a) and (b) 
    represent five contours at the values ($0, \pm 5, \pm 10$) of $Q_{12}$ and 
    $Q_{21}$, respectively, as shown on the figures. In the white areas in (a) and 
    (b), the coupling coefficients $Q_{12}$ (for low $k_2$) and $Q_{21}$ (for low 
    $k_1$), respectively, assume very high values that have been omitted for 
    clarity.
    The half-difference $\delta$ (c) and the half-sum $V$ (d) of the group 
    velocities are also shown on the $k_1 - k_2$ plane. The black curves on the
    surface of (c) and (d) are contours at $\delta =0$ and $V=0.5$, respectively. 
    }
    \label{fig2}
\end{figure}
%****************************************************************************fig

Once the above compatibility condition has been applied, the equations arising 
in second order can be solved, for various harmonic amplitudes. One thus obtains 
coupled solutions (for the two wavepackets) in the form of a superposition of 
first and second harmonics and combinations (i.e. sum and difference) thereof. 
(Details are provided in the Appendix.)

Proceeding to third order ($\sim \varepsilon^3$), once again, a  pair of 
compatibility conditions must be imposed for annihilation of secular terms. 
One is thus led to a pair of coupled NLS equations in the form
\begin{eqnarray}
\label{eq35}
   i \left( \frac{\partial \Psi_1}{\partial t_2} 
   +v_{g,1} \frac{\partial \Psi_1}{\partial x_2} \right)
   +P_1 \frac{\partial^2 \Psi_1}{\partial x_1^2}
   +\left(Q_{11} |\Psi_1|^2 +Q_{12} |\Psi_2|^2 \right) \Psi_1 =0, 
\\
\label{eq36}
   i \left( \frac{\partial \Psi_2}{\partial t_2} 
   +v_{g,2} \frac{\partial \Psi_2}{\partial x_2} \right)
   +P_2 \frac{\partial^2 \Psi_2}{\partial x_1^2}
   +\left(Q_{21} |\Psi_1|^2 +Q_{22} |\Psi_2|^2 \right) \Psi_2 =0, 
\end{eqnarray}
where
\begin{equation}
\label{eq37}
   P_j =-\frac{3}{2} c_1 \frac{k_j}{ (k_j^2 +c_1)^{5/2} }, ~~~~~~(j=1,2)
\end{equation} 
are the dispersion coefficients and 
\begin{equation}
\label{eq38}
    Q_{jj} = \frac{\omega_j}{2 k_j^2} \tilde{Q}_{jj}, \qquad
    Q_{12} = \frac{\omega_1}{2 k_1^2} \tilde{Q}_{12}, \qquad
    Q_{21} = \frac{\omega_2}{2 k_2^2} \tilde{Q}_{21}
\end{equation} 
are the self modulation and cross-coupling coefficients, respectively. 
(Expressions for the tilded quantities are provided in the Appendix.)
Note that $Q_{12}$ and $Q_{21}$ exchange expressions, upon permuting $k_1$ and 
$k_2$ (and $\omega_1$ and $\omega_2$), as intuitively expected, upon a simple 
symmetry argument. 
%*****************************************************************************80

As we shall see below, the coefficients $P_j$ and $Q_{ij}$ in Eqs. (\ref{eq35}) 
and (\ref{eq36}) 
do not exhibit any particular symmetry for arbitrary values of the wavenumbers 
$k_1$ and $k_2$, as e.g. in the famous Manakov model \cite{Manakov1974} which 
was proven to be completely integrable. During the last decades, other different 
variants of CNLS systems of equations were investigated in various contexts and 
were proven to be integrable \cite{Mikhailov1981,Zen1998,Gerdjikov2004,Wang2010}. 
Note that our CNLS system recovers all of these systems in appropriate limits. 
However, as stated above, it is a-symmetric in its general form.

%****************************************************************************fig
In Fig. \ref{fig1}(a) the frequency dispersion $\omega_1$, the group velocity 
$v_{g1}$, and the dispersion coefficient $P_1$ are all plotted as a function of 
the wavenumber of the first carrier wave $k_1$. The corresponding nonlinearity 
coefficient $Q_{11}$ and the ratio $Q_{11}/P_1$ are shown also as a function of 
$k_1$ in Fig. \ref{fig1}(b). The corresponding quantities for the second carrier 
wave (i.e., $\omega_2$, $v_{g2}$, $Q_{22}$, etc.) as functions of the 
carrier wavenumber $k_2$ have exactly the same form (upon shifting $k_1$ to $k_2$). 
Note that, for the cold-ion fluid model considered here, the dispersion 
coefficients $P_j$ ($j = 1, 2$) are both negative for any $k_j$, in fact 
converging asymptotically to zero from below for large $k_j$. 
(This fact is expected to change if one considers a warm ion fluid.) On the other 
hand, the nonlinearity coefficients $Q_{jj}$ can be either positive or negative, 
depending on the value of $k_j$: they both exhibit one real root at $k_j =1.47$ as 
expected from previous works \cite{Kakutani1974} (and refs. therein), so that 
$Q_{jj} > 0$ for $k_j < 1.47$ and $Q_{jj} < 0$ otherwise.

The coupling coefficients $Q_{12}$ and $Q_{21}$ given in Eqs. (\ref{eq38}) 
clearly depend on both $k_1$ and $k_2$, and they exhibit a reflection symmetry 
around $k_1 =k_2$, as intuitively expected. Two-dimensional maps for $Q_{12}$ 
and $Q_{21}$ on the $k_1 - k_2$ wavenumber plane are shown in Figs. \ref{fig2}(a) 
and \ref{fig2}(b), respectively. The half-difference and the half-sum of the 
group velocities $\delta$ and $V$, respectively, are shown on the $k_1 - k_2$ 
plane in Figs. \ref{fig2}(c) and \ref{fig2}(d).

%*****************************************************************************80
\subsection{Transformation to the regular CNLS equations form}

Equations (\ref{eq22}) and  (\ref{eq23}) can be transformed to the most common 
form of CNLS equations with a change of independent variables, followed by a 
transformation. Let us introduce the new variables $\xi$ and $\tau$ through 
(drop the subscripts in $x$ and $t$)
\begin{equation}
\label{eq_t01}
   \xi = x -v t, \qquad \tau =t,
\end{equation}
where $v =( v_{g,1} +v_{g,2} )/2$, is the half-sum of the group velocities 
$v_{g,1}$ and $v_{g,2}$. Using $v$ and the half-difference of the group 
velocities $\delta =( v_{g,1} -v_{g,2} )/2$ we have that 
\begin{equation}
\label{eq_t04}
   v_{g,1} = v +\delta, \qquad v_{g,2} = v -\delta.
\end{equation}
By applying the change of variables to Eqs.  (\ref{eq22}) and  (\ref{eq23}) 
we get 
\begin{eqnarray}
\label{eq_t05}
   i \left( \frac{\partial \Psi_1}{\partial \tau} 
   +\delta \frac{\partial \Psi_1}{\partial \xi} \right)
   +P_1 \frac{\partial^2 \Psi_1}{\partial \xi^2}
   +\left\{Q_{11} |\Psi_1|^2 +Q_{12} |\Psi_2|^2 \right\} \Psi_1 =0, 
\\
\label{eq_t06}
   i \left( \frac{\partial \Psi_2}{\partial \tau} 
   -\delta \frac{\partial \Psi_2}{\partial \xi} \right)
   +P_2 \frac{\partial^2 \Psi_2}{\partial \xi^2}
   +\left\{Q_{21} |\Psi_1|^2 +Q_{22} |\Psi_2|^2 \right\} \Psi_2 =0.
\end{eqnarray}
In writing the latter equations, it is implicitly understood that the 
differentiations involved in the dispersive terms are with respect to the 
variable $\xi_1$, while the higher-order differentiation in the beginning of 
the LHS are with respect to $\xi_2$. The subscripts are nonetheless omitted, 
for simplicity, as they will not affect the final result. Next, we apply to 
Eqs. (\ref{eq_t05}) and  (\ref{eq_t06}) the following transformation
\begin{eqnarray}
\label{eq_t07}
   \Psi_1 = \bar{\Psi}_1 \exp\left[ i \left( \frac{\delta^2}{4 P_1} \tau 
      -\frac{\delta}{2 P_1} \xi \right) \right],
\\
\label{eq_t08}
   \Psi_2 = \bar{\Psi}_2 \exp\left[ i \left( \frac{\delta^2}{4 P_2} \tau 
    +\frac{\delta}{2 P_2} \xi \right) \right],
\end{eqnarray}
where $\bar{\Psi}_1$ and $\bar{\Psi}_2$ are complex functions of the 
new variables $\xi$ and $\tau$, to get after some straightforward algebra:
\begin{eqnarray}
\label{eq_t09}
   i \frac{\partial \bar{\Psi}_1}{\partial \tau} 
   +P_1 \frac{\partial^2 \bar{\Psi}_1}{\partial \xi^2}
   +\left\{Q_{11} |\bar{\Psi}_1|^2 +Q_{12} |\bar{\Psi}_2|^2 \right\} \bar{\Psi}_1 =0, 
\\
\label{eq_t10}
   i \frac{\partial \bar{\Psi}_2}{\partial \tau} 
   +P_2 \frac{\partial^2 \bar{\Psi}_2}{\partial \xi^2}
   +\left\{Q_{21} |\bar{\Psi}_1|^2 +Q_{22} |\bar{\Psi}_2|^2 \right\} \bar{\Psi}_2 =0.
\end{eqnarray}

Formally equivalent CNLS equations to those in Eqs. (\ref{eq35})-(\ref{eq36}), 
or to those in (\ref{eq_t09})-(\ref{eq_t10}), have been obtained earlier, for 
different plasma fluid models 
\cite{Spatschek1978,Som1979,McKinstrie1989,McKinstrie1990,Luther1990,Luther1992,Singh2013,Borhanian2017}, 
as mentioned in the introduction, and also in several other physical contexts, 
e.g., in pulse propagation in water waves \cite{He2022}, in deep water wave 
propagation \cite{Ablowitz2015}, in the study of asymmetric coupled wavefunctions 
\cite{Kourakis2006}, in soliton propagation in left-handed transmission lines 
\cite{Veldes2013}, in left-handed metamaterials \cite{Lazarides2005}, in the 
study of pulses of different polarizations in anisotropic dispersive media 
\cite{Tyutin2022}, in pulse propagation in optical nonlinear media 
\cite{Kivshar1993}, and in randomly birefringent optical fibers 
\cite{Frisquet2016}, among others. Theoretical and computational studies of 
envelope solitons in coupled NLS models have been presented in Ref. 
\cite{Kevrekidis2016}, while a brief overview of various localized solutions in 
the Manakov model is given in Ref. \cite{Priya2013}. 
Note that when $P_j =1$ and $Q_{11} =Q_{21} =-Q_{22} =-Q_{12}$ our system reduces 
to the one treated in Refs. \cite{Zakharov1982,Kanna2006}.

%*****************************************************************************80
\paragraph{An alternative form of the CNLS system.}  \ \ \ 
It is worth pointing out that the number of nontrivial parameters in the CNLS 
equations Eqs. (\ref{eq_t09}) and (\ref{eq_t10}) can be reduced by rescaling  
the dependent and the independent variables. One may follow the procedure in 
\cite{Veldes2013,Tan-Boyd2000} by assuming $Q_{11}, Q_{22} > 0$ and $P_1 \ne 0$,  
and simultaneously applying the transformations
\begin{equation}
\label{eq_t11}
   \hat{\Psi}_j =\bar{\Psi}_j \sqrt{Q_{jj}}, \qquad
   \zeta =\xi/\sqrt{|P_1|},
\end{equation}
to Eqs. (\ref{eq_t09}) and (\ref{eq_t10}). One thus obtains
\begin{eqnarray}
\label{eq_t15}
   i \frac{\partial \hat{\Psi}_1}{\partial \tau} 
   +s \frac{\partial^2 \hat{\Psi}_1}{\partial \zeta^2}
   +\left\{ |\hat{\Psi}_1|^2 +\mu_{12} |\hat{\Psi}_2|^2 \right\} \hat{\Psi}_1 =0, 
\\
\label{eq_t16}
   i \frac{\partial \hat{\Psi}_2}{\partial \tau} 
   +p \frac{\partial^2 \hat{\Psi}_2}{\partial \zeta^2}
   +\left\{ |\hat{\Psi}_2|^2 +\mu_{21} |\hat{\Psi}_1|^2 \right\} \hat{\Psi}_2 =0,
\end{eqnarray}
where 
\begin{equation}
\label{eq_t14}
  \mu_{12} =\frac{Q_{12}}{Q_{22}}, \qquad
  \mu_{21} =\frac{Q_{21}}{Q_{11}}, \qquad
  p =\frac{P_2}{|P_1|}, \qquad s= \frac{P_1}{|P_1|} = \pm 1
\end{equation}
and the number of nontrivial parameters has thus been reduced to three 
\cite{Tan-Boyd2000, Veldes2013}, or rather four (including $s$, i.e. the sign of 
$P_1$). Do not forget that this conclusion comes under the assumption that the 
dispersion coefficient $P_1$ is finite, i.e. non-zero, and that the nonlinearity 
coefficients $Q_{11}$ and $Q_{22}$ are all positive. Recall that $P_1 < 0 $, 
actually, in the cold-ion model considered here, hence $s =-1$ here.

%*****************************************************************************80
\section{Modulational stability analysis: analytical framework}

The modulational stability analysis for two co-propagating wavepackets in the 
plasma described by the fluid equations (\ref{eq01}) may now be performed, 
relying on the CNLS equations (\ref{eq_t05}) and (\ref{eq_t06}) following the 
procedure of Refs. \cite{Kourakis2006,Borhanian2017}. For this purpose we first 
seek a plane-wave solution of the form 
\begin{equation}
\label{mi01}
   \Psi_j =\Psi_{j,0} \, e^{i \tilde{\omega}_j \tau},
\end{equation}
where $\Psi_{j,0}$ ($j=1,2$) is a constant real amplitude and 
$\tilde{\omega}_j$ is a real frequency correction to be determined.

By inserting Eq. (\ref{mi01}) into Eqs. (\ref{eq_t05}) and (\ref{eq_t06}), we get
\begin{equation}
\label{mi02}
   \tilde{\omega}_1 =Q_{11} \Psi_{1,0}^2 +Q_{12} \Psi_{2,0}^2, \qquad {\rm and} 
   \qquad \tilde{\omega}_2 =Q_{21} \Psi_{1,0}^2 +Q_{22} \Psi_{2,0}^2 , .
\end{equation}
Eq. (\ref{mi01}), with the frequency correction(s) given by Eq. (\ref{mi02}), 
provide a linear coupled-mode solution of the CNLS system (a coupled Stokes 
wave pair, actually, in the hydrodynamic picture) wherein the propagation of 
waves is governed by Eqs. (\ref{eq_t05}) and (\ref{eq_t06}). In order to 
address the stability of the above plane-wave solutions against a small 
perturbation, we take 
\begin{equation}
\label{mi03}
  \Psi_j =( \Psi_{j,0} +\varepsilon_j ) \, e^{i \tilde{\omega}_j \tau},
\end{equation}
where $\varepsilon_j$ represents a complex perturbation on the amplitudes 
of the two waves ($|\varepsilon_j| \ll \Psi_{j,0}$). By inserting the perturbed 
amplitudes $\Psi_j$ into Eqs. (\ref{eq_t05}) and (\ref{eq_t06}), we obtain 
\begin{eqnarray}
\label{mi04}
 i \left(\frac{\partial \varepsilon_1}{\partial \tau} 
    +\delta \frac{\partial \varepsilon_1}{\partial \xi} \right)
 +P_1 \frac{\partial^2 \varepsilon_1}{\partial \xi^2} 
    +Q_{11} \Psi_{1,0}^2 (\varepsilon_1 +\varepsilon_1^\star) 
    +Q_{12} {\Psi}_{1,0} \Psi_{2,0} (\varepsilon_2 +\varepsilon_2^\star) =0, \\
 \label{mi05}
 i \left(\frac{\partial \varepsilon_2}{\partial \tau} 
    -\delta \frac{\partial \varepsilon_2}{\partial \xi} \right)
 +P_2 \frac{\partial^2 \varepsilon_2}{\partial \xi^2} 
    +Q_{22} \Psi_{2,0}^2 (\varepsilon_2 +\varepsilon_2^\star) 
    +Q_{21} \Psi_{2,0} \Psi_{1,0} (\varepsilon_1 +\varepsilon_1^\star) =0.
 \end{eqnarray}  
Setting $\varepsilon_j =g_j +i h_j$, where $g_j$ and $h_j$ are real functions, 
then subsequently substituting into Eqs. (\ref{mi04}) and (\ref{mi05}) and eventually 
eliminating the imaginary parts $h_j$ from the resulting equations, we obtain for 
the real parts: 
\begin{eqnarray}
\label{mi06}
\left\{ \left(\frac{\partial}{\partial \tau} 
   +\delta \frac{\partial}{\partial \xi} \right)^2 
   +P_1 \left( P_1 \frac{\partial^2}{\partial \xi^2} 
   +2 Q_{11} \Psi_{1,0}^2 \right) \frac{\partial^2}{\partial \xi^2} \right\} g_1
   +2 P_1 Q_{12} \Psi_{2,0} \Psi_{1,0}  \frac{\partial^2}{\partial \xi^2} g_2 =0, 
 \\
 \label{mi07}
 \left\{ \left(\frac{\partial}{\partial \tau} 
   -\delta \frac{\partial}{\partial \xi} \right)^2 
   +P_2 \left( P_2 \frac{\partial^2}{\partial \xi^2} 
   +2 Q_{22} \Psi_{2,0}^2 \right) \frac{\partial^2}{\partial \xi^2} \right\} g_2
   +2 P_2 Q_{21} \Psi_{1,0} \Psi_{2,0}  \frac{\partial^2}{\partial \xi^2} g_1 =0 \, .
 \end{eqnarray}  
Eventually, by setting (i.e. considering harmonic amplitude perturbations) 
\begin{equation}
\label{mi08}
   g_j =g_{j,0} \, e^{i (K \xi -\Omega \tau)} +c.c. ,
\end{equation}
where $c.c.$ denotes the complex conjugate,
and then substituting into the earlier equations, we obtain after some 
algebra a {\em compatibility condition}
\begin{equation}
\label{mi09}
   \left[ (\Omega -K \delta)^2 -\Omega_1^2 \right] 
   \left[ (\Omega +K \delta)^2 -\Omega_2^2 \right] =\Omega_c^4,
\end{equation}
where
\begin{equation}
\label{mi10}
  \Omega_1^2 =P_1 K^2 \left( P_1 K^2 -2 Q_{11} \Psi_{1,0}^2 \right), ~~~
  \Omega_2^2 =P_2 K^2 \left( P_2 K^2 -2 Q_{22} \Psi_{2,0}^2 \right), ~~~
  \Omega_c^4 =4 P_1 P_2 Q_{12} Q_{21} \Psi_{1,0}^2 \Psi_{2,0}^2 K^4.      
\end{equation}
This is essentially a dispersion relation for the amplitude perturbation 
considered above. So long as \emph{all} solutions for $\Omega$ are real, the 
amplitude perturbation will be stable to external disturbances, e.g. noise. 
Conversely, for instability to set in, at least one (of the four) solutions 
of the quartic polynomial Eq. (\ref{mi09}) in $\Omega$ should have a positive 
imaginary part. 

The growth rate of the modulational instability is then defined as
\begin{equation}
\label{mi11}
   \Gamma = {\rm max}\{Im(\Omega)\} ,        
\end{equation}
i.e. is given by the maximum value of the imaginary part of the solution. 
(Note that either two complex conjugate solutions will exist, or otherwise four 
-- i.e. two pairs of complex conjugates --  in which case the maximum value will 
dominate. Negative values of the imaginary part will lead to decaying modes.)  

\medskip

%*****************************************************************************80
\textit{Relevance with Space plasmas.   }
Evidence of MI processes in space was found since the very early stages of space 
missions \cite{Ergun1991,Bonnell1997,Weatherall1997,Ruderman2010}. On several 
occasions, the theory of MI is capable of explaining the generation of modulated 
wave envelopes commonly observed in space plasmas, as well as solitary waves and 
freak or rogue waves. Such processes, in particular, can explain the detection 
of electric field wave envelops observed in the terrestrial auroral zone 
\cite{Ergun1991,Bonnell1997}. In Ref.  \cite{Ergun1991}, the authors reported 
direct evidence of nearly $100\%$ electric field amplitude modulation of Langmuir 
waves (see their Figure 2) observed in the auroral ionosphere. These data are 
consistent with a transverse MI, whose characteristic frequencies are near the 
observed modulational frequencies (obtained by independent measurements). 
In Ref. \cite{Bonnell1997}, several hundreds of large amplitude and narrow-band 
modulated Langmuir waves (observed by Freja satellite and SCIFER sounding rocket 
in the terrestrial auroral zone) were used to estimate their modulation frequencies. 
It was shown that lower hybrid waves in the cold electrostatic limit can produce 
the estimated modulation frequencies (i.e., several tens of kilohertz). 
In Ref. \cite{Weatherall1997}, the author analyzed a model of coupled mode 
equations describing plasma wave turbulence in the strongly magnetized pair plasma 
of a pulsar polar cap. Numerical solutions of the model equations reveal the 
development of turbulence, whose onset involves a secondary wave-wave 
(modulational) instability. In Ref. \cite{Ruderman2010}, the possibility of 
generating large-amplitude and short-lived wavepackets from small-amplitude 
initial perturbations in various plasmas is discussed. The author presents a 
figure (Figure 1 in his article) for the hour averages of the magnetic field 
strength in the heliosheath as a function of time, in which the magnitude of the 
field exhibits many spike-like dips (magnetic holes) which are characterized as 
freak plasma waves. The author shows that for a particular wave mode, the 
ion-sound wave in plasma with negative ions described by Gardner equation, a 
modulationally unstable initial condition leads through MI to the generation of 
freak or rogue ion-acoustic waves. From the discussion of these works it can be 
inferred that the mechanism of the MI in various space plasmas is very important 
and relevant for the generation of modulated wavepackets (envelops) as well as 
large amplitude solitary waves (freak or rogue waves), among others, which have 
been observed by space missions. In other words, the generation of these 
structures seem to be a consequence of MI processes which may occur frequently in
plasmas.

%*****************************************************************************80
\section{Modulational instability: parametric analysis based on numerical results}
%*****************************************************************************80

We have computed the growth rate $\Gamma$ numerically, by finding the roots of 
the polynomial in $\Omega$ based on Eq. (\ref{mi09}), and subsequently
identifying the one with the highest imaginary part. That highest imaginary part 
was then plotted as a function of the wavenumber of the perturbation $K$ and the 
wavenumber of the second carrier wave $k_2$, considering three different values 
of $k_1$ and -- independently -- for two different pairs of values of $A_{1,0}$ 
and $A_{2,0}$. 

\begin{figure}[htp]
    \centering
    \includegraphics[width=7.5cm]{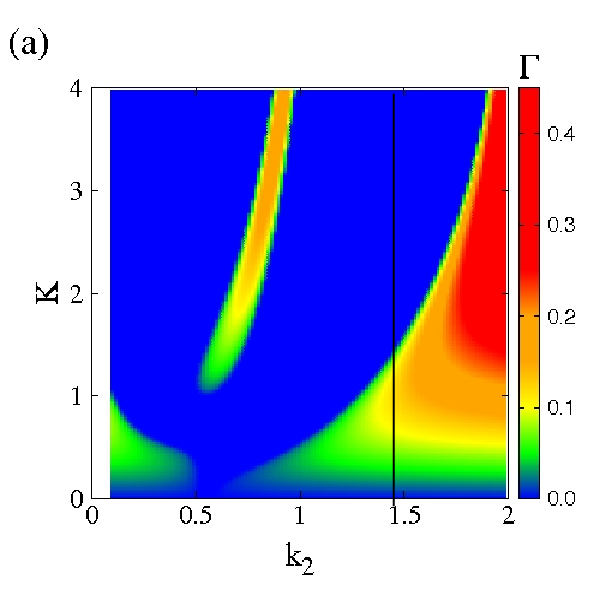}
    \includegraphics[width=7.5cm]{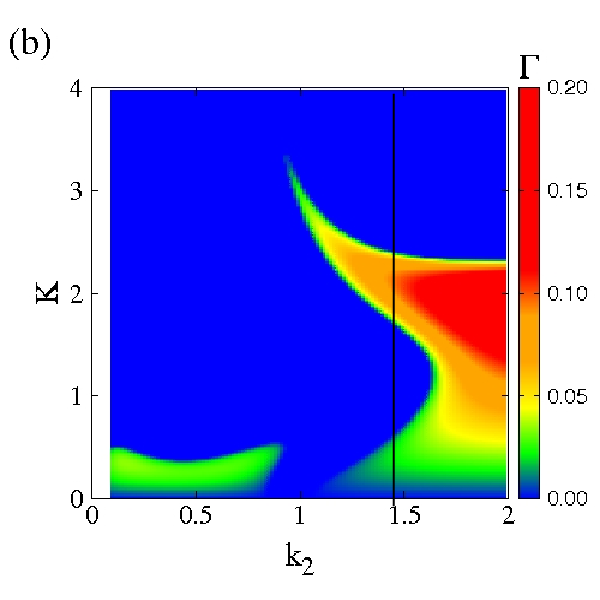} \\
    \includegraphics[width=7.5cm]{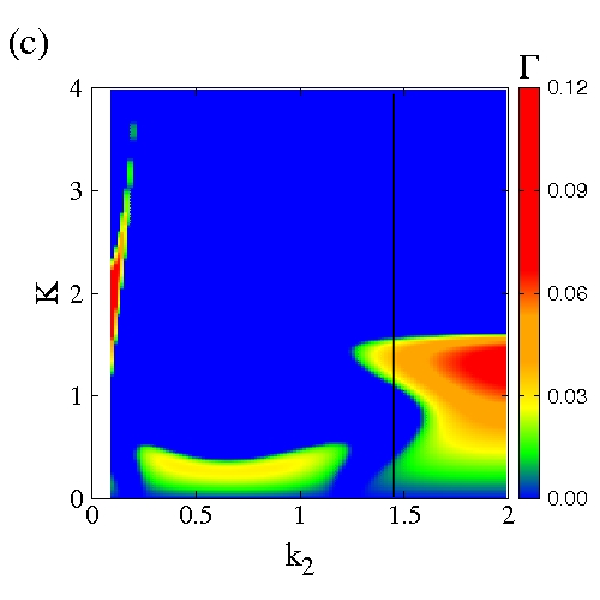}
    \includegraphics[width=7.5cm]{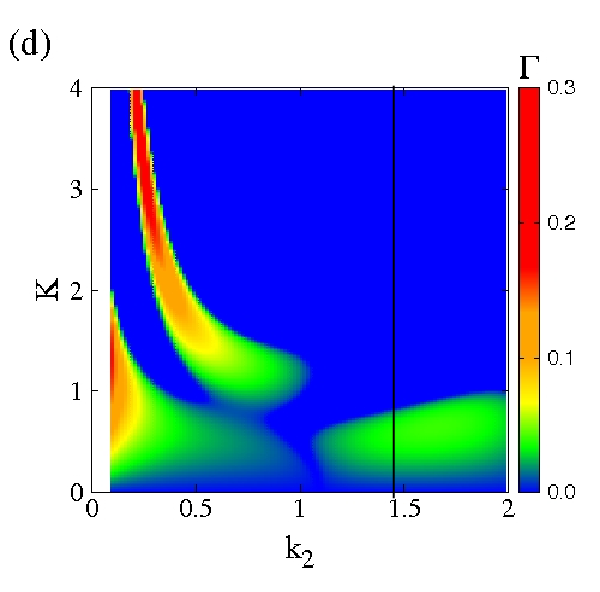}
    \caption{The growth rate $\Gamma$ as a function of the wavenumber of the 
             second carrier wave $k_2$ and the wavenumber of the perturbation $K$ 
             for $A_{1,0} =A_{2,0} =0.15$, and 
    (a) $k_1 =0.1$,
    (b) $k_1 =0.4$,
    (c) $k_1 =0.7$,
    (d) $k_1 =1.6$.
    The black vertical line is located at $k_2 =1.47$ and separates areas on
    the plane with positive (left from the line) and negative (right from the 
    line) $Q_{22}$.
    }
    \label{fig3}
\end{figure}

\begin{figure}[htp]
    \centering
    \includegraphics[width=7.5cm]{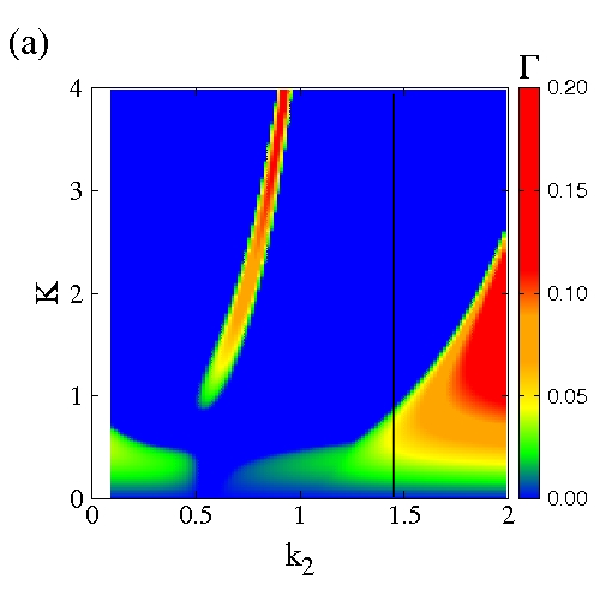}
    \includegraphics[width=7.5cm]{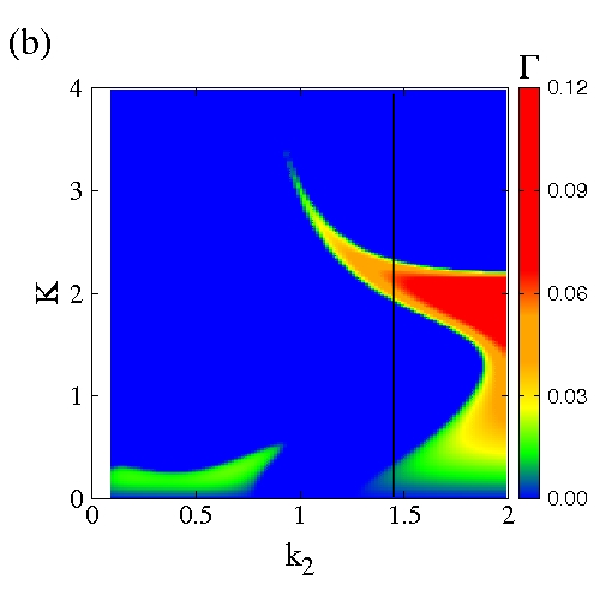} \\
    \includegraphics[width=7.5cm]{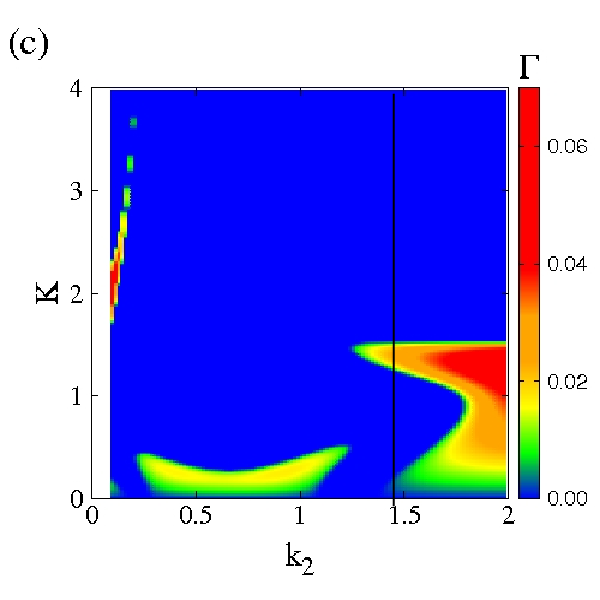}
    \includegraphics[width=7.5cm]{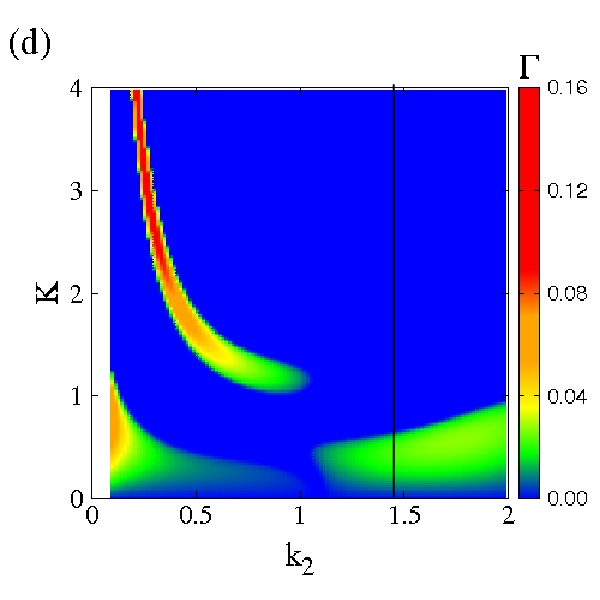}
    \caption{The growth rate $\Gamma$ as a function of the wavenumber of the 
            second carrier wave $k_2$ and the wavenumber of the perturbation $K$ 
            for $A_{1,0} =0.0707$, $A_{2,0} =0.2$, and 
    (a) $k_1 =0.1$,
    (b) $k_1 =0.4$, 
    (c) $k_1 =0.7$,
    (d) $k_1 =1.6$.
    The black vertical line is located at $k_2 =1.47$ and separates areas on
    the plane with positive (left from the line) and negative (right from the 
    line) $Q_{22}$.
    }
    \label{fig4}
\end{figure}

In Figs. \ref{fig3}(a), (b), (c) and (d) (for equal amplitudes, 
$A_{1,0} =A_{2,0} =0.15$), the four values of the carrier wavenumber $k_1$ are 
respectively $0.1$, $0.4$, $0.7$ and $1.6$. (Note that the first three were chosen 
in the modulationally stable region $k_1 < 1.47$, whereas the latter was chosen 
from the intrinsically unstable range $k_1 > 1.47$.) The figures reveal in all 
cases shown a rather complex instability pattern. The analogous patterns for  
$A_{1,0} =0.0707$ and $A_{2,0} =0.2$, shown in Figs. \ref{fig4}, reveal that the
growth rate is generally lower in this case (for unequal amplitudes $A_{1,0}$ and 
$A_{2,0}$) while at the same time the $K$ intervals (``windows") of instability 
are narrower. Note that the two pairs of amplitudes were chosen so that the norm 
$A_{1,0}^2 +A_{2,0}^2$ (qualitatively representing the total electrostatic energy 
injected in the system) remains the same. For the values of $k_1$ and $k_2$ used 
to obtained the results in the first three panels (a)-(c) of both Figs. \ref{fig3} 
and \ref{fig4}, the decoupled Eqs. (\ref{eq_t05}) and (\ref{eq_t06}) resulting by 
setting $Q_{12} =Q_{21} =0$ are both stable against periodic perturbations since 
$P_j\, Q_{jj} < 0$. (Recall that $P_j$ is always negative in the considered model 
and that $Q_{jj}$ is positive for $k_j < 1.47$.) In contrast to this picture, the 
fourth panel (d), in both of these figures, corresponds to the unstable case 
$k_1 > 1.47$, for the first wave. On the other hand, the second wavepacket may 
either be (independently) stable (for $k_2 < 1.47$) or unstable (for $k_2 > 1.47$), 
i.e. in the absence of its sister wave 1. Implicitly, all four combinations --- 
say ${\rm sgn}(P_j Q_{jj}) \in (+/+, +/-, -/+, -/-)$ --- in the combined stability 
profile of the wave pair have thus been taken into consideration (albeit with 
arbitrary, sample parameter values).

Significant qualitative conclusions may be drawn based on critical inspection of 
our indicative numerical plots. If one draws a vertical line at $k_2 = 1.47$, in 
any of the (8) plots presented in Figs. \ref{fig3} and \ref{fig4}, 
the area on the left (right) of that vertical line will correspond to a region 
where the 2nd wave is intrinsically unstable (stable). As expected, in panels (a)-(c) 
of both of Figs. \ref{fig3} and \ref{fig4}, the map-profile is darker (blue color 
is dominant) on the left half of the plots: this reveals that, when both waves -- in 
the absence of each other's sister wave -- are modulationally stable, their coupling 
may lead to the system being destabilized, yet this is rather limited to regions of 
large perturbation wavenumber $K$ (see the small islands top left, in Figures 
\ref{fig3}(c), \ref{fig4}(a) and \ref{fig4}(c)). The resulting instability (due to 
the nonlinear coupling between two otherwise stable wavepackets) is more intense 
in the case of unequal amplitudes: cf. \ref{fig4}(a) with \ref{fig3}(a)), for 
strong $K$ and low $k_2$.

\begin{figure}[htp]
    \centering
    \includegraphics[width=15cm]{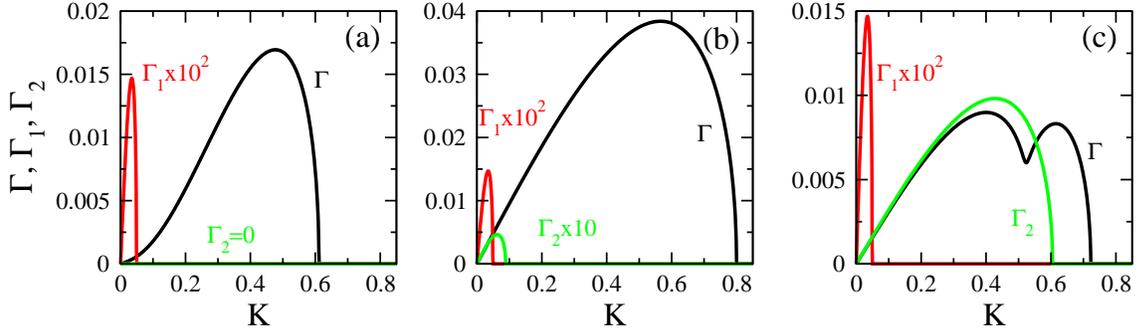}
    \caption{The growth rate $\Gamma$ and the the growth rates $\Gamma_1$ and 
            $\Gamma_2$ of the first and second decoupled Eqs. (\ref{eq_t05}) and 
            (\ref{eq_t06}) as a function of the wavenumber of the perturbation 
            $K$, for $k_1 =1.48$ ($Q_{11} < 0$), $A_{1,0} =A_{2,0} =0.15$, and 
    (a) $k_2 =1$, 
    (b) $k_2 =1.5$,
    (c) $k_2 =2$.
    }
    \label{fig5}
\end{figure}

On the other hand, it is interesting to observe the last sub-panel (d) in either 
of Figs. \ref{fig3} and \ref{fig4}. In this one, one of the waves (wave 1) is 
intrinsically unstable. Therefore, in the right half of sub-panels \ref{fig3}(d) 
and \ref{fig4}(d) (were both $k_1$ and $k_2$ are above 1.47), both waves are 
modulationally unstable (independently from one another). Still,  
somehow counter-intuitively, it appears that the coupling results in them being 
stabilized somehow, for large $K$ (note the blue-colored stability region in the 
top right quadrant in both these plots). 
However, by comparing the growth rate $\Gamma$ in Figs. \ref{fig3}(d) and 
\ref{fig4}(d) with the corresponding rates $\Gamma_1$ and $\Gamma_2$ of the 
two single NLS equations resulting from the decoupled CNLS system, we observe that 
$\Gamma$ is always larger in magnitude and and that the instability interval in 
$K$ is much larger than the corresponding ones of $\Gamma_1$ and $\Gamma_2$.
An illustrative example of this situation can be seen in  Fig. \ref{fig5}(b) below.

On the other hand, the left half of these sub-plots corresponds to a situation 
where one (1) of the waves (2) is stable, while the other one (1) is unstable. 
As seen in the other plots too, this combination leads to an unstable pair of 
wavepackets. 

%****************************************************************************fig
%****************************************************************************fig

%****************************************************************************fig
%****************************************************************************fig

As discussed above, when either one of the $k_j$s or both $k_j$s is (are) 
greater than $1.47$, the system of the CNLS Eqs. (\ref{eq_t05}) and 
(\ref{eq_t06}) is again modulationally unstable for $K$ values within certain 
intervals (instability windows). An illustrative example is shown in Fig. 
\ref{fig5} for $k_1 =1.48$ and $k_2 =1$ (a), $k_2 =1.5$ (b), and $k_2 =2$ (c). 
In this figure, the growth rate $\Gamma$ of the system of Eqs. 
(\ref{eq_t05})-(\ref{eq_t06}) is compared with the growth rates $\Gamma_1$ and 
$\Gamma_2$, obtained respectively from the first and second decoupled Eqs. 
(\ref{eq_t05}) and (\ref{eq_t06}) upon ``switching off" the coupling, setting 
$Q_{ij}=0$ for $i \ne j$ for a minute --- or, more accurately, upon setting 
$k_2 =0$ in Eq. (\ref{eq_t05}) and $k_1 =0$ in Eq. (\ref{eq_t05}). For those 
decoupled equations, in particular, by using the same modulational instability 
analysis as it was used in the beginning of this section, we find that the 
growth rates $\Gamma_1$ and $\Gamma_2$ are given by the magnitude of the 
imaginary part of the perturbation frequency $\Omega$ obtained from the 
equations
\begin{equation}
\label{mi12}
    \Omega =+\delta K \pm \Omega_1, \qquad \Omega =-\delta K \pm \Omega_2,
\end{equation}
for the first and the second decoupled NLS equation, respectively.
The results shown in Fig. \ref{fig5} reveal that the growth rate $\Gamma$ of 
the system of coupled NLS equations is always much larger that the growth rates
$\Gamma_1$ and $\Gamma_2$ of the decoupled equations. Moreover, in the CNLS
equations, the instability occurs in much larger intervals of the wavenumber of
the perturbation $K$.  
As a consequence, asides the qualitative effect of the coupling (discussed in 
the previous paragraph), its \emph{quantitative} impact is dramatic, in that 
wave-wave coupling results in an increased growth rate and an enhanced 
instability window.

%****************************************************************************fig
%****************************************************************************fig

\subsection{Special case: decoupled second wave, ($Q_{21} =0$)} 
In the general case, the choice of a pair of wavenumbers $(k_1, k_2)$ leads 
generally to different values for $Q_{ij}$ and $P_j$. As mentioned earlier, 
the nonlinearity coefficients $Q_{jj}$ change their sign, from positive to 
negative, as the wavenumber $k_j$ exceeds $1.47$. Moreover, as one may have 
notice from Figs. \ref{fig2}(a) and (b), $Q_{12}$ or $Q_{21}$ (but not both) 
can become zero on the particular curve on the $k_1 - k_2$ plane. Let us choose 
a pair of $(k_1, k_2)$ values for which, say $Q_{21}$ is zero. In that case, 
Eq. (\ref{eq_t06}) is not affected by Eq. (\ref{eq_t05}), although the latter 
is still affected by the former. In this interesting situation, the perturbation 
dispersion relation reduces into two independent quadratic equations. Each of 
them is the dispersion relation for an NLS equation which is uncoupled from the 
other NLS equation. The roots of the dispersion relation in this case are then 
the same as those in Eq. (\ref{mi12}). Thus, the condition for stability is the 
same as that for the two decoupled NLS equations, i.e., $P_1 Q_{11} < 0$ and 
$P_2 Q_{22} < 0$.

\begin{figure}[htp]
    \centering
    \includegraphics[width=15cm]{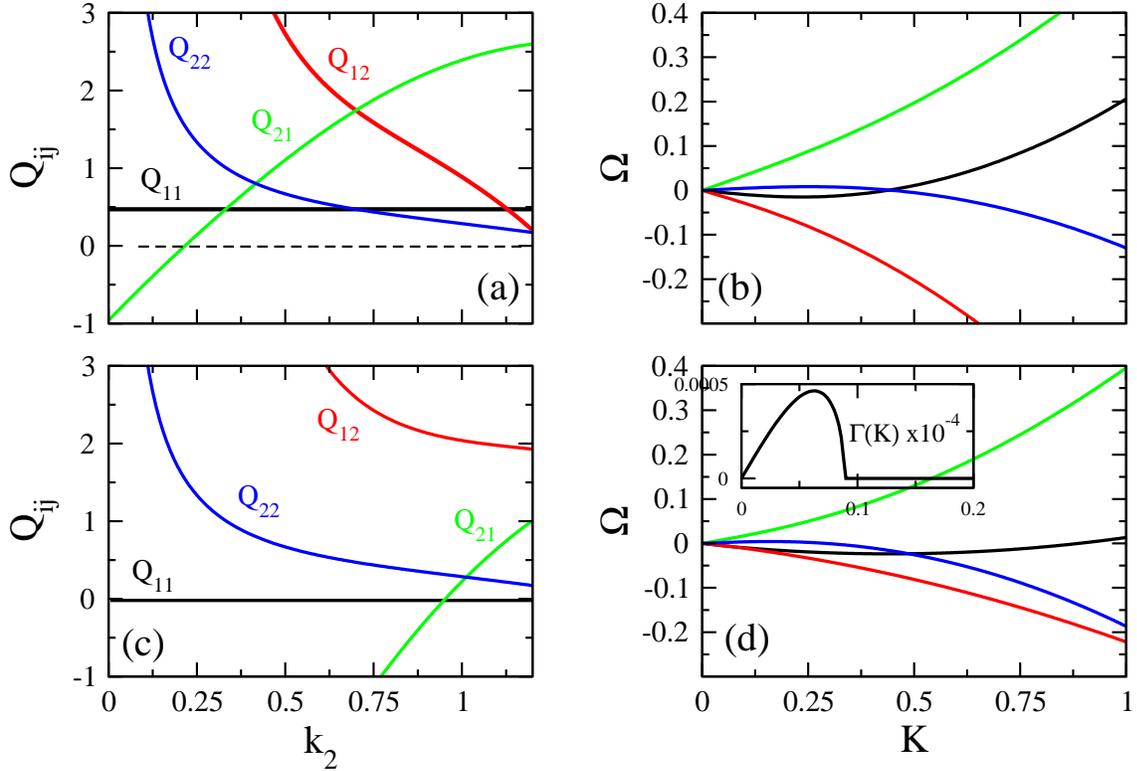}
    \caption{
    (a) The coupling and nonlinearity coefficients $Q_{ij}$ ($i,j =1,2$) as a
    function of the wavenumber $k_2$ of the second carrier wave for $k_1 =0.7$.
    The coupling coefficient $Q{21}$ becomes zero at $k_2 \simeq 0.2160$.  
    (b) The real parts of the perturbation frequency dispersion branches 
    $\Omega =\Omega(K)$ for $k_1 =0.7$, $k_2 \simeq 0.2160$, and 
    $A{1,0} =A_{2,0} =0.15$. The corresponding imaginary parts are zero.
    (c) As in (a) for $k_1 =1.5$. $Q{21}$ becomes zero at $k_2 \simeq 0.9529$.
    (d) As in (b) for $k_1 =1.5$, $k_2 \simeq 0.9529$, and 
    $A_{1,0} =A_{2,0} =0.15$.
    Inset: The imaginary part of the branch of $\Omega$ whose real part is 
    shown in black color, that coincides with the corresponding growth rate
    $\Gamma$.
    }
    \label{fig6}
\end{figure}

In Figs. \ref{fig6}(a) and \ref{fig6}(c), the $Q_{ij}$s are plotted as a function
of $k_2$ for fixed $k_1 =0.7$ and $1.5$, respectively. For these parameter values,
the coupling coefficient $Q_{21}$ becomes zero at $k_2 \simeq 0.2160$ and $0.9529$,
respectively. Subsequently, for the pairs of $(k_1, k_2)$ where $Q_{21} =0$, i.e.,
for $(k_1, k_2) =(0.7, 0.2160)$ and $(k_1, k_2) =(1.5, 0.9529)$, the perturbation
frequency dispersions are plotted in Figs. \ref{fig6}(b) and \ref{fig6}(d),
respectively. In Fig. \ref{fig6}(b) all frequency dispersion branches are real,
since both $k_1, k_2 < 1.47$ and $P_1 Q_{11} < 0$, $P_2 Q_{22} < 0$, thus the 
CNLS system is stable in this case. In Fig. \ref{fig6}(d), however, two of the 
frequency branches are complex in a particular interval of the perturbation 
frequency $K$, since in this case $k_2 < 1.47 < k_1$ and 
$P_2 Q_{22} < 0 < P_1 Q_{11}$. The branch with the positive imaginary part
actually determines the (in)stability interval in $K$ of the whole system as 
well as the corresponding growth rate. That imaginary part is plotted in the 
inset of Fig. \ref{fig6}(d), where we observe that the system is unstable for 
$K$ in $(0, 0.09)$.

%****************************************************************************fig
%****************************************************************************fig

%*****************************************************************************80
\section{Conclusions}
%*****************************************************************************80

A set of coupled nonlinear Schr{\"o}dinger (CNLS)  equations was derived for a 
pair of co-propagating (and interacting) electrostatic wavepackets, governed by 
a plasma fluid model comprising a cold inertial ion fluid evolving against a 
thermalized electron background. A multiple-scale technique has been adopted, 
similar to the well known Newell method in nonlinear optics. 

The dispersion, nonlinearity, and coupling coefficients $P_j$ and $Q_{ij}$ 
($j=1,2$) depend on the carrier wavenumbers $k_1$ and $k_2$ (both assumed 
arbitrary) of the two interacting wavepackets, and in general they differ
from each other. The exact mathematical expressions for those coefficients 
have been analytically obtained, and in fact depend parametrically on the 
plasma parameters too. Although $P_j$ for the particular model considered 
here assumes negative values that tend to zero for large $k_j$, the coefficients 
$Q_{jj}$ can be either positive or negative, actually becoming zero at $k_j =1.47$ 
as discussed earlier. The coupling coefficients $Q_{12}$ and $Q_{21}$, which 
depend both on $k_1$ and $k_2$, assume their values from a large interval of 
positive and negative real numbers. All this prescribes a multifaceted dynamical 
profile with strong interconnection among the various physical parameters.
 
A compatibility condition was derived through modulational stability analysis, 
in the form of a fourth degree polynomial in the frequency of the perturbation. 
By numerically finding the roots/solutions of the compatibility condition and 
identifying the one with the largest magnitude of the imaginary part (growth 
rate), modulationally stable and unstable areas on the $k_2 - K$ plane have been 
identified. The instability growth rate, mapped on that plane, reveals that 
modulational instability occurs in areas of the $k_2 - K$ plane (say, for fixed 
$k_1$), actually for all four combinations of the sign of the product $P_j Q_{jj}$ 
of the decoupled equations, i.e. obtained by appropriately selecting 
the wavenumbers $k_1$ and $k_2$. (Recall that stability imposes $P_j Q_{jj} < 0$ 
in the single NLS case.)

Modulational instability in fact may occur even if both waves of the decoupled 
system are stable, since their strong nonlinear coupling may potentially 
destabilize the system in regions with large perturbation wavenumbers $K$. The 
instability that results from the nonlinear coupling between  the two wavepackets 
is more intense in the case of equal amplitudes of the two waves; instability 
areas also occupy larger areas (``windows") on the $k_2 - K$ plane in the latter 
case. When one or both the waves of the decoupled system are unstable, their 
nonlinear coupling leads again to an unstable pair of wavepackets as shown 
earlier. 

A comparison of the growth rates of the two waves of the decoupled system with 
that of the waves in the coupled one reveals that the latter is always larger 
than the former; furthermore, the unstable regions on the $k_2 - K$ plane are 
larger than those of the waves in the decoupled system.

Note that the modulational instability of the nonlinear co-propagating waves 
may lead dynamically to the formation of localized electrostatic waveforms of 
solitonic type (envelope solitons), i.e. essentially components of a vector 
soliton, while the plasma system remains physically stable. This is a commonly 
met situation in complex dynamical systems in which a solution loses its 
stability for a particular parameter set while at the same time a stable solution 
emerges.

The electron component in this work is assumed to follow the Maxwell-Boltzmann 
distribution, for simplicity. This ``simple" model successfully captures the 
salient features of modulational interactions between electrostatic plasma 
wavepackets as regards their envelope dynamics. As a straightforward generalization, 
our model can be extended to cover a non-Maxwellian particle distribution, 
such as e.g., a kappa distribution,
a common occurrence in Space plasmas. Work in this direction is in progress; 
preliminary results suggesting a strong dependence of the modulational profile 
on the electron statistics, were recently presented at a conference 
\cite{ICPPNikos}. Once completed, this work will form the focus of a forthcoming 
article.

%*****************************************************************************80
\section*{Acknowledgements}
Authors IK and NL gratefully acknowledge financial support from Khalifa 
University of Science and Technology, Abu Dhabi, United Arab Emirates, via the 
project CIRA-2021-064 (8474000412).
IK gratefully acknowledges financial support from KU via the project FSU-2021-012 
(8474000352) as well as from KU Space and Planetary Science Center, 
%Abu Dhabi,  United Arab Emirates,  
via grant No. KU-SPSC-8474000336. 
The contributions (analytical work) by two of us (AJ and GV) to this project were 
carried out during their respective  research visit in 2022, funded by an ADEK 
(Abu Dhabi Education and Knowledge Council, currently ASPIRE-UAE) AARE 
(Abu Dhabi Award of Research Excellence) grant (AARE18-179). 
Hospitality from the host during those visits is gratefully acknowledged.
Author AJ acknowledges support by Tamkeen under the NYU (Nea York University 
Abu Dhabi) Research Institute grant CG008. 
This work was completed while one of us (IK) was a Visiting Researcher at the 
National and Kapodistrian University of Athens (Greece) during a sabbatical 
leave;  % hosted by Prof. D.J.  Frantzeskakis; 
the hospitality of the host is warmly acknowledged. 

%*****************************************************************************80

%*****************************************************************************80
\appendix
%*****************************************************************************80
\section{Detailed algebraic procedure}

\subsection{Multiple scales perturbation technique}

We introduce fast and slow variables, viz. $x_n =\varepsilon^n x$ and 
$t_n =\varepsilon^n t$ (for $n = 0,1,2,3,...$), where $\varepsilon \ll 1$ is a 
small real constant, so that the partial derivatives in Eqs. (\ref{eq01}) are 
approximated as
\begin{eqnarray}
\label{eq04}
   \frac{\partial}{\partial t} \rightarrow  \frac{\partial}{\partial t_0} 
                                +\varepsilon \frac{\partial}{\partial t_1} 
                                +\varepsilon^2 \frac{\partial}{\partial t_2} 
                                + \varepsilon^3 \frac{\partial}{\partial t_3} 
                                +\cdots,
   \qquad
%\label{eq05}   
   \frac{\partial}{\partial x} \rightarrow  \frac{\partial}{\partial x_0} 
                                +\varepsilon \frac{\partial}{\partial x_1} 
                                +\varepsilon^2 \frac{\partial}{\partial x_2} 
                                +\varepsilon^3 \frac{\partial}{\partial x_3} 
                                +\cdots ,
\end{eqnarray}
hence
\begin{eqnarray}
\label{eq06}
   \frac{\partial^2}{\partial x^2} \rightarrow  \frac{\partial^2}{\partial x_0^2} 
  +\varepsilon \left( 2 \frac{\partial^2}{\partial x_0 \partial x_1} \right) 
  +\varepsilon^2 \left( \frac{\partial^2}{\partial x_1^2} 
   +2 \frac{\partial^2}{\partial x_0 \partial x_2} \right)
  + \varepsilon^3 \left( 2 \frac{\partial^2}{\partial x_0 \partial x_3} 
   +2 \frac{\partial^2}{\partial x_1 \partial x_2} \right)
  +\cdots,
\end{eqnarray} 
The dependent variables $n$, $u$, and $\phi$ can be expanded in powers of 
$\varepsilon$ as
\begin{equation}
\label{eq07}
   n =1 +\varepsilon n_1 +\varepsilon^2 n_2 +\varepsilon^3 n_3 +\cdots, \qquad
   u =\varepsilon u_1 +\varepsilon^2 u_2 +\varepsilon^3 u_3 +\cdots, \qquad 
   \phi =\varepsilon \phi_1 +\varepsilon^2 \phi_2 +\varepsilon^3 \phi_3 +\cdots.
\end{equation}
The expansions Eq. (\ref{eq07}) are substituted into Eqs. (\ref{eq01}), and using 
the approximations (\ref{eq04}) and (\ref{eq06}) for the derivatives we 
obtain sets of equations at different orders in $\varepsilon$, proceeding up to 
order $\varepsilon^3$.

Specifically, the equations in the first order ($\propto \varepsilon^1$) read
 \begin{eqnarray}
\label{eq_i01}
   \frac{\partial n_1}{\partial t_0} +\frac{\partial (u_1)}{\partial x_0} =0, 
   \qquad
%\label{eq_i02}
   \frac{\partial u_1}{\partial t_0} +\frac{\partial \phi_1}{\partial x_0} =0,
   \qquad
%\label{eq_i03}
   \frac{\partial^2 \phi_1}{\partial x_0^2} +n_1 -c_1 \phi_1 =0,
\end{eqnarray}
where $x_0 =x$ and $t_0 =t$. In these equations, the following {\em Ansatz} is 
introduced
 \begin{eqnarray}
\label{eq_i04}
  S_1 =S_{1,1}^{(-1)} e^{-i \theta_1} +S_{1,1}^{(1)} e^{+i \theta_1} 
         +S_{1,2}^{(-1)} e^{-i \theta_2} +S_{1,2}^{(1)} e^{+i \theta_2},
\end{eqnarray}
where $S_1 = \{n_1, u_1, \phi_1\}$ is the state vector (triplet) to 1st order, 
and the phases are given by $\theta_j =k_j x -\omega_j t$, with $k_j$ and 
$\omega_j$ being the wavevector and the corresponding frequency of the $j-$th 
wave; note that reality of the state variables imposes 
$S_{1,j}^{(-1)} = (S_{1,j}^{(1)})^*$ (for $j=1,2$), where the star (*) denotes 
the complex conjugate. Note that the derivatives with respect to the fast 
variables $t=t_0$ and $x=x_0$ that appear in Eqs. (\ref{eq_i01}) do not act on 
the amplitudes $S_{1,j}^{(\pm 1)}$, since the latter does not depend on $t$ 
and $x$. After some straightforward calculations we get the equations
\begin{eqnarray}
\label{eq_i09}
   -\omega_j n_{1,j}^{(1)} +k_j u_{1,j}^{(1)} =0, 
   \qquad
   -\omega_j u_{1,j}^{(1)} +k_j \phi_{1,j}^{(1)} =0,
   \qquad
  +n_{1,j}^{(1)} -(k_j^2 +c_1) \phi_{1,j}^{(1)} =0,
\end{eqnarray}
%*****************************************************************************80
with $j = 1, 2$. Clearly, the equations for $j=1$ are not coupled to those with 
$j=2$ and thus the two waves do not interact to each other at this order of 
approximation. As a result, two uncoupled $3\times 3$ systems with unknowns 
($n_{1,1}^{(1)}$, $u_{1,1}^{(1)}$, $\phi_{1,1}^{(1)}$), and ($n_{1,2}^{(1)}$, 
$u_{1,2}^{(1)}$, $\phi_{1,2}^{(1)}$), respectively, are obtained of the form    
\begin{eqnarray}
\label{eq_i15}
\begin{bmatrix}
   -i \omega_j & +i k_j      &  0  \\
    0          & -i \omega_j & +i k_j  \\
    1          & 0           & -(k_j^2 +c_1) 
\end{bmatrix}
\begin{bmatrix}
   n_{1,j}^{(1)}     \\
   u_{1,j}^{(1)}     \\
  \phi_{1,j}^{(1)}       
\end{bmatrix}
 =
\begin{bmatrix}
  0    \\
  0    \\
  0     
\end{bmatrix}
,
\end{eqnarray}
with $j=1,2$. For these systems to have a non-trivial solution, it is required 
that their determinants should vanish, i.e., 
\begin{eqnarray}
\label{eq_i16}
D_j =
\begin{vmatrix}
   -\omega_j & +k_j       &  0  \\
     0       & -\omega_1  & +k_j  \\
     1       & 0          & -(k_j^2 +c_1) 
\end{vmatrix}
=0,
\end{eqnarray}
which provides the linear frequency dispersion relations and the corresponding 
group velocities as
\begin{equation}
\label{eq_i17}
   \omega_j =\frac{k_j}{\sqrt{k_j^2 +c_1}}, 
   \qquad 
   v_{g,j} =\frac{\partial \omega_j}{\partial k_j} =\frac{c_1}{(k_j^2 +c_1)^{3/2}}.
\end{equation}
(Note that only the positive branch for both $\omega$ and $k$ is considered.)
In the following, the wavenumbers $k_j$ and their corresponding 
frequency $\omega_j$ will appear in the derived mathematical expressions. This 
is done only for convenience and for ease in the presentation and wherever it 
occurs, it is implied that $\omega_j$ is provided by Eq. (12) as a function of $k_j$.

The systems of Eqs. (\ref{eq_i15}) can now be solved in terms of the variables 
$\phi_{1,j}^{(1)} =\Psi_j$ ($j=1,2$) to give
\begin{equation}
\label{eq_i17.2}
    n_{1,j}^{(1)} =\left(\frac{k_j}{\omega_j}\right)^2 \Psi_j, 
    \qquad
    u_{1,j}^{(1)} =\frac{k_j}{\omega_j} \Psi_j,
    \qquad 
    \phi_{1,j}^{(1)} =\Psi_j.
\end{equation}

The equations in the second order in $\varepsilon$ read
 \begin{eqnarray}
\label{eq11}
   \frac{\partial n_2}{\partial t_0} +\frac{\partial (u_2)}{\partial x_0} ={\cal F}_1,
   \qquad
   \frac{\partial u_2}{\partial t_0} +\frac{\partial \phi_2}{\partial x_0} ={\cal F}_2,
   \qquad
   \frac{\partial^2 \phi_2}{\partial x_0^2} -c_1 \phi_2 +n_2 ={\cal F}_3,
\end{eqnarray}
where
\begin{eqnarray}
\label{eq12}
   {\cal F}_1 = -\frac{\partial (n_1)}{\partial t_1} 
                -\frac{\partial (u_1)}{\partial x_1} 
                -\frac{\partial (n_1 u_1)}{\partial x_0},
   \qquad
   {\cal F}_2 =-\frac{\partial u_1}{\partial t_1} 
               -\frac{\partial \phi_1}{\partial x_1} 
               -u_1 \frac{\partial u_1}{\partial x_0},
   \qquad
   {\cal F}_3 =+c_2 \phi_1^2 -2 \frac{\partial^2 \phi_1}{\partial x_0 \partial x_1}.
\end{eqnarray}
The right-hand sides of Eqs. (\ref{eq11}), i.e., the function ${\cal F}_1$, 
${\cal F}_2$, and ${\cal F}_1$, are known functions of the leading order 
potential amplitudes $\Psi_j$ that result from substitution of Eqs. (\ref{eq_i17.2}) 
into Eqs. (\ref{eq12}). Then, in order to solve Eqs. (\ref{eq11}), we use the 
following {\em Ansatz}
\begin{equation}
\label{eq12.2}
   S_2 =S_{2}^{(0)} +\left\{ S_{2,1}^{(1)} e^{i \theta_1} +S_{2,1}^{(2)} e^{2 i \theta_1}
                 +S_{2,2}^{(1)} e^{i \theta_2} +S_{2,2}^{(2)} e^{2 i \theta_2}
                 +S_{2,+}^{(1)} e^{i (\theta_1 +\theta_2)} +S_{2,-}^{(1)} 
                    e^{2 i (\theta_1 -\theta_2)} +C. C.
                    \right\},
\end{equation}
where $S = $ any of $n$, $u$ or $\phi$, and $C.C.$ denotes the complex conjugate 
of the expression within the curly brackets. By substitution of Eqs. (\ref{eq12.2}) 
into Eqs. (\ref{eq11}), we obtain the system of equations 
\begin{eqnarray}
\label{eq12.3}
   -\omega_j n_{2,j}^{(1)} +k_j u_{2,j}^{(1)} =
    -\left( \frac{k_j}{\omega_j} \right)^2 \frac{\partial \Psi_j}{\partial t_1} 
    -\frac{k_j}{\omega_j} \frac{\partial \Psi_j}{\partial x_1} =\mu_{1,j}, 
   \qquad
   -\omega_j u_{2,j}^{(1)} +k_j \phi_{2,j}^{(1)} =
    -\frac{k_j}{\omega_j} \frac{\partial \Psi_j}{\partial t_1}
     -\frac{\partial \Psi_j}{\partial x_1} =\mu_{2,j}, 
\nonumber \\
   -(k_j^2 +c_1) \phi_{2,j}^{(1)} +n_{2,j}^{(1)} =
    -2 i k_j \frac{\partial\Psi_j}{\partial x_1} =\mu_{3,j},
\end{eqnarray}
resulting from all terms proportional to $e^{i \theta_1}$ and $e^{i \theta_2}$, 
respectively, which will be discussed later. Evaluating the action of the 
derivatives, one obtains an algebraic system in the form 
\begin{equation}
\label{eq12.4}
   -2 \omega_j n_{2,j}^{(2)} +2 k_j u_{2,j}^{(2)} =-2 k_j n_{1,j}^{(1)} u_{1,j}^{(1)}, 
   \qquad
   -2 \omega_j u_{2,j}^{(2)} +2 k_j \phi_{2,j}^{(2)} =-k_j u_{1,j}^{(1)}{^2},
   \qquad
  -(4 k_j^2 +c_1) \phi_{2,j}^{(2)} +n_{2,j}^{(2)}=c_2 \Psi_j^2,
\end{equation}
for $j=1$ and $2$, resulting from all terms proportional to $e^{2 i \theta_1}$ 
and $e^{2 i \theta_2}$, respectively, with solutions
\begin{equation}
\label{eq12.5}
  n_{2,j}^{(2)} =C_{n,2,j}^{(2)} \Psi_j^2, \qquad 
  u_{2,j}^{(2)} =C_{u,2,j}^{(2)} \Psi_j^2, \qquad 
  \phi_{2,j}^{(2)} =C_{\phi,2,j}^{(2)} \Psi_j^2,
\end{equation}
where the coefficients are given in the Appendix, and 
\begin{eqnarray}
\label{eq12.6}
   -(\omega_1 +\omega_2) n_{2,+}^{(1)} +(k_1 +k_2) u_{2,+}^{(1)} 
   =-(k_1 +k_2) \frac{k_1 k_2}{\omega_1 \omega_2} 
       \left( \frac{k_1}{\omega_1} +\frac{k_2}{\omega_2} \right) \Psi_1 \Psi_2 =
       f_1^{(+)} \Psi_1 \Psi_2, 
   \\ 
   \label{eq12.7}
   -(\omega_1 +\omega_2) u_{2,+}^{(1)} +(k_1 +k_2) \phi_{2,+}^{(1)} 
   =-(k_1 +k_2) \frac{k_1 k_2}{\omega_1 \omega_2} \Psi_1 \Psi_2 =
   f_2^{(+)} \Psi_1 \Psi_2,
   \\
   \label{eq12.8}
  -\left[(k_1+k_2)^2 +c_1 \right] \phi_{2,+}^{(2)} +n_{2,+}^{(1)}=
   +2 c_2 \Psi_1 \Psi_2 =f_3^{(+)} \Psi_1 \Psi_2,
\end{eqnarray}
and 
\begin{eqnarray}
\label{eq12.09}
   -(\omega_1 -\omega_2) n_{2,-}^{(1)} +(k_1 -k_2) u_{2,-}^{(1)} 
   =-(k_1 -k_2) \frac{k_1 k_2}{\omega_1 \omega_2} 
       \left( \frac{k_1}{\omega_1} +\frac{k_2}{\omega_2} \right) 
        \Psi_1 \Psi_2^\star =f_1^{(-)} \Psi_1 \Psi_2^\star, 
   \\ \label{eq12.10}
   -(\omega_1 -\omega_2) u_{2,-}^{(1)} +(k_1 -k_2) \phi_{2,-}^{(1)} 
   =-(k_1 -k_2) \frac{k_1 k_2}{\omega_1 \omega_2} \Psi_1 \Psi_2^\star =
   f_2^{(-)} \Psi_1 \Psi_2^\star,
   \\
   \label{eq12.11}
  -\left[(k_1 -k_2)^2 +c_1 \right] \phi_{2,-}^{(2)} +n_{2,-}^{(1)}=
   +2 c_2 \Psi_1 \Psi_2^\star =f_3^{(-)} \Psi_1 \Psi_2^\star \, .
\end{eqnarray}
The last two $3 \times 3$ systems result from the terms proportional to 
$e^{i (\theta_1 +\theta_2)}$ and $e^{i (\theta_2 -\theta_2)}$, respectively. 
Their solution reads 
\begin{eqnarray}
\label{eq12.13}
   n_{2,+}^{(1)} =C_{n,2,+}^{(1)} \Psi_1 \Psi_2, \qquad u_{2,+}^{(1)} =
   C_{u,2,+}^{(1)} \Psi_1 \Psi_2, \qquad
   \phi_{2,+}^{(1)} =C_{\phi,2,+}^{(1)} \Psi_1 \Psi_2,
\\
\label{eq12.14}
   n_{2,-}^{(1)} =C_{n,2,-}^{(1)} \Psi_1 \Psi_2^\star, \qquad u_{2,-}^{(1)} =
   C_{u,2,-}^{(1)} \Psi_1 \Psi_2^\star, \qquad
   \phi_{2,-}^{(1)} =C_{\phi,2,-}^{(1)} \Psi_1 \Psi_2^\star,
\end{eqnarray}
where again the coefficients $C_{n,2,\pm}^{(1)}$, $C_{u,2,\pm}^{(1)}$, and 
$C_{\phi,2,\pm}^{(1)}$ are given in the Appendix.

Note that, at order $\varepsilon^2$, we also obtain an equation that contains 
the zeroth-order (``constant'') terms, i.e.  terms which do not depend on the 
zeroth order independent variables $x=x_0$ and $t=t_0$, as
\begin{equation}
\label{eq12.15}
   -c_1 \phi_2^{(0)} +n_2^{(0)} =+2 c_2 \left( |\Psi_1|^2 +|\Psi_2|^2 \right),
\end{equation} 
which will be used in the next subsection.

%*****************************************************************************80
Before moving on to the third order equations  in $\varepsilon$, we return to 
the discussion of the solution of Eqs. (\ref{eq12.3}). The latter, which 
actually consist of two $3\times 3$ uncoupled systems for the variables 
$\{ n_{2,1}^{(1)}, u_{2,1}^{(1)}, \phi_{2,1}^{(1)} \}$ and 
$\{ n_{2,2}^{(1)}, u_{2,2}^{(1)}, \phi_{2,2}^{(1)} \}$, can be written 
in matrix form as
\begin{eqnarray}
\label{eq13}
\begin{bmatrix}
   -\omega_j   & +k_j        &  0  \\
    0          & -\omega_j   & +k_j  \\
    1          & 0           & -(k_j^2 +c_1) 
\end{bmatrix}
\begin{bmatrix}
   n_{2,j}^{(1)}     \\
   u_{2,j}^{(1)}     \\
  \phi_{2,j}^{(1)}       
\end{bmatrix}
 =
\begin{bmatrix}
  \mu_{1,j}    \\
  \mu_{2,j}    \\
  \mu_{3,j}     
\end{bmatrix}
,
\end{eqnarray}
where $\mu_{m,j}$ ($m=1,2,3$ and $j=1,2$) are defined in Eq. (\ref{eq12.3}). 
Note that the determinants of the two inhomogeneous systems of Eqs. (\ref{eq13}) 
are zero, since this was a compatibility requirement imposed in the first order, 
that led to the frequency dispersion relations. Thus, in order to find nontrivial 
solutions for the two inhomogeneous systems of Eqs. (\ref{eq13}), it must be 
imposed that the following determinants 
\begin{eqnarray}
\label{eq14}
D_j' =
\begin{vmatrix}
   \mu_{1,j}    & +k_j       &  0    \\
   \mu_{2,j}    & -\omega_1  & +k_j  \\
   \mu_{3,j}    & 0          & -(k_j^2 +c_1) 
\end{vmatrix}
=0 
\end{eqnarray}
should also vanish. These determinants result from the replacement of the 
first column of the determinant $D_j$ with the vector 
${\bf \mu}_j =[ \mu_{1,j} \, \mu_{2,j} \, \mu_{3,j} ]^T$. After some algebra, 
we obtain that in order for $D_j'$ to vanish, the conditions 
\begin{equation}
\label{eq15}
   \frac{\partial \Psi_j}{\partial t_1} =-v_{g,j} \frac{\partial \Psi_j}{\partial x_1}
\end{equation} 
should hold (compatibility conditions). Recall that the group velocity, defined 
earlier, naturally arose through the compatibility conditions appearing in this 
order. Physically speaking, this means that the wavepacket amplitudes move at 
the group velocity, as expected. This is a well known result, related with the 
so called Newell technique in nonlinear optics, that has been discussed e.g. in 
Ref. \cite{Infeld-Rowlands}.
%*****************************************************************************80

%*****************************************************************************80
\subsection{Derivation of the coupled CNLS equations}

The equations in order $\varepsilon^3$ are
\begin{eqnarray}
\label{eq16}
   \frac{\partial n_3}{\partial t_0} +\frac{\partial (u_3)}{\partial x_0} ={\cal G}_1,
   \qquad
   \frac{\partial u_3}{\partial t_0} +\frac{\partial \phi_3}{\partial x_0} ={\cal G}_2,
   \qquad
   \frac{\partial^2 \phi_3}{\partial x_0^2} -c_1 \phi_3 +n_3 ={\cal G}_3,
\end{eqnarray}
where
\begin{eqnarray}
\label{eq17}
  {\cal G}_1 =-\frac{\partial (n_2)}{\partial t_1} -\frac{\partial (u_2)}{\partial x_1}
   -\frac{\partial (n_1)}{\partial t_2} -\frac{\partial (u_1)}{\partial x_2}
   -\frac{\partial (n_1 u_2 +n_2 u_1)}{\partial x_0} -\frac{\partial (n_1 u_1)}{\partial x_1},
  \\
\label{eq18}
   {\cal G}_2 =-\frac{\partial u_2}{\partial t_1} -\frac{\partial \phi_2}{\partial x_1} 
   -\frac{\partial u_1}{\partial t_2} -\frac{\partial \phi_1}{\partial x_2}
   -\frac{\partial u_1 u_2}{\partial x_0} -u_1 \frac{\partial u_1}{\partial x_1},
   \\
\label{eq19} 
   {\cal G}_3 =-2 \frac{\partial^2 \phi_2}{\partial x_0 \partial x_1} 
   -2 \frac{\partial^2 \phi_1}{\partial x_0 \partial x_2}
   -\frac{\partial^2 \phi_1}{\partial x_1^2} +2 c_2 \phi_1 \phi_2 +c_3 \phi_1^3.
\end{eqnarray}
We may now substitute into Eqs. (\ref{eq16})-(\ref{eq19}) the following {\em Ansatz}
\begin{eqnarray}
\label{eq20}
   S_3 =S_{3}^{(0)} +\left\{ S_{3,1}^{(1)} e^{i \theta_1} +S_{3,1}^{(2)} e^{2 i \theta_1} +S_{3,1}^{(3)} e^{3 i \theta_1}
                            +S_{3,2}^{(1)} e^{i \theta_2} +S_{3,2}^{(2)} e^{2 i \theta_2} +S_{3,2}^{(3)} e^{3 i \theta_2}
                            \right.
                            \nonumber \\
                            \left.
                            +S_{3,+}^{(1)} e^{i (\theta_1 +\theta_2)} +S_{3,-}^{(1)} e^{2 i (\theta_1 -\theta_2)} 
                            +S_{3,+12}^{(1)} e^{i (\theta_1 +2\theta_2)} +S_{3,-12}^{(1)} e^{i (\theta_1 -2\theta_2)}
                            \right.
                            \nonumber \\
                            \left.
                            +S_{3,+21}^{(1)} e^{i (2\theta_1 +\theta_2)} +S_{3,-21}^{(1)} e^{i (2\theta_1 -\theta_2)} +C. C.
                            \right\},
\end{eqnarray}
where $S =n, u$ or $\phi$ and $C. C.$ denotes the complex conjugate of the 
entire expression within the curly brackets. For the $\ell=0$ mode, we obtain 
two equations for the variables $n_2^{(0)}$ and $u_2^{(0)}$, which will be 
combined with Eq. (\ref{eq12.15}) to provide a $3\times 3$ system with the 
unknowns $n_2^{(0)}$, $u_2^{(0)}$, and $\phi_2^{(0)}$, viz. 
\begin{eqnarray}
\label{eq21}
   \frac{\partial n_2^{(0)}}{\partial t_1} +\frac{\partial u_2^{(0)}}{\partial x_1} 
   =-2 \frac{\partial}{\partial x_1} \left[ \left( \frac{k_1}{\omega_1} \right)^3 |\Psi_1|^2 
                                           +\left( \frac{k_2}{\omega_2} \right)^3 |\Psi_2|^2 \right],
   \\
   \label{eq22}
   \frac{\partial u_2^{(0)} }{\partial t_1} +\frac{\partial \phi_2^{(0)}}{\partial x_1} 
   =-\frac{\partial}{\partial x_1} \left[ \left( \frac{k_1}{\omega_1} \right)^2 |\Psi_1|^2
                                         +\left( \frac{k_2}{\omega_2} \right)^2 |\Psi_2|^2 \right]
   \\
   \label{eq23}
   n_2^{(0)} =c_1 \phi_2^{(0)} +2 c_2 \left( |\Psi_1|^2 +|\Psi_2|^2 \right),
\end{eqnarray}
where the relations (\ref{eq_i17.2}) were also used. To solve Eqs. 
(\ref{eq21})-(\ref{eq23}), we shall introduce the {\em Ansatz} 
\begin{equation}
\label{eq24}
  \phi_2^{(0)} =C_{\phi,2,1}^{(0)} |\Psi_1|^2 +C_{\phi,2,2}^{(0)} |\Psi_2|^2,
  \qquad
  u_2^{(0)} =C_{u,2,1}^{(0)} |\Psi_1|^2 +C_{u,2,2}^{(0)} |\Psi_2|^2,
  \qquad
  n_2^{(0)} =C_{n,2,1}^{(0)} |\Psi_1|^2 +C_{n,2,2}^{(0)} |\Psi_2|^2,
\end{equation} 
into Eqs. (\ref{eq21})-(\ref{eq23}) and use Eq. (\ref{eq15}) to determine 
the coefficients $C_{\phi,2,j}$, $C_{u,2,j}$ and $C_{n,2,j}$, whose 
analytical expressions are given in the Appendix. Finally, the equations 
for the (1st harmonic) mode $\ell =1$ (to third order in $\varepsilon$) 
read
\begin{equation}
\label{eq25}
  -i \omega_j n_{3,j}^{(1)} +i k_j u_{3,j}^{(1)} ={\cal R}_{1,j},
  \qquad
  -i \omega_j u_{3,j}^{(1)} +i k_j \phi_{3,j}^{(1)} ={\cal R}_{2,j},
  \qquad
  -\left( k_j^2 +c_1 \right) \phi_{3,j}^{(1)} +n_{3,j}^{(1)} ={\cal R}_{3,j},
\end{equation}
where
\begin{eqnarray}
\label{eq26}
  {\cal R}_{1,j} =-\left( \frac{\partial n_{2,j}^{(1)} }{\partial t_1} 
                  +\frac{\partial u_{2,j}^{(1)} }{\partial x_1} 
                         +\frac{\partial n_{1,j}^{(1)} }{\partial t_2} 
                         +\frac{\partial u_{1,j}^{(1)} }{\partial x_2}
  \right)
  -i k_j \left( n_{1,j}^{(-1)} u_{2,j}^{(2)} +n_{1,j}^{(1)} u_{2}^{(0)} 
     +n_{1,k_j}^{(-1)} u_{2,p}^{(1)} +n_{1,k_j}^{(1)} u_{2,m}^{(\ell_j)}
  \right)
  \nonumber \\
  -i k_j \left( u_{1,j}^{(-1)} n_{2,j}^{(2)} +u_{1,j}^{(1)} n_{2}^{(0)} 
     +u_{1,k_j}^{(-1)} n_{2,p}^{(1)} +u_{1,k_j}^{(1)} n_{2,m}^{(\ell_j)}
  \right),
  \\
  \label{eq27}
  {\cal R}_{2,j} =-\left( \frac{\partial u_{2,j}^{(1)} }{\partial t_1} 
                         +\frac{\partial \phi_{2,j}^{(1)} }{\partial x_1} 
                         +\frac{\partial u_{1,j}^{(1)} }{\partial t_2} 
                         +\frac{\partial \Psi_j }{\partial x_2}
                         \right)
  -i k_j \left( u_{1,j}^{(-1)} u_{2,j}^{(2)} +u_{1,j}^{(1)} u_{2}^{(0)} 
     +u_{1,k_j}^{(-1)} u_{2,p}^{(1)} +u_{1,k_j}^{(1)} u_{2,m}^{(\ell_j)}
  \right),  
  \\
   \label{eq28}
   {\cal R}_{3,j} =-2 i k_j \left( \frac{\partial \phi_{2,j}^{(1)} }{\partial x_1} 
   +\frac{\partial \Psi_j }{\partial x_2} \right)
   -\frac{\partial^2 \Psi_j }{\partial x_1^2}
   +2 c_2 \left( \Psi_j^\star \phi_{2,j}^{(2)} +\Psi_j \phi_{2}^{(0)} 
   +\Psi_{k_j}^\star \phi_{2,p}^{(1)} +\Psi_{k_j} \phi_{2,m}^{(\ell_j)} \right)
   \nonumber \\
   +3 c_3 \Psi_j \left( |\Psi_j|^2 +2 |\Psi_{k_j}|^2 \right),
\end{eqnarray}
where $j=1,2$, and $k_j$, $\ell_j$ defined as $\ell_1 =1$, $\ell_2 =-1$, 
$k_1 =2$, and $k_2 =1$.
By combining properly Eqs. (\ref{eq25}), we arrive to a new equation 
from which the variables $n_{3,j}^{(1)}$, $u_{3,j}^{(1)}$ and 
$\phi_{3,j}^{(1)}$ have been eliminated (essentially, a condition for 
annihilation of secular terms) in the form
\begin{equation}
\label{eq29}
  -i \omega_j {\cal R}_{1,j} -i k_j {\cal R}_{2,j} +\omega_j^2 {\cal R}_{2,j} =0,
\end{equation}
where the expressions for ${\cal R}_{m,j}$ ($m=1,2,3$ and $j=1,2$) were given by 
Eqs. (\ref{eq26})-(\ref{eq28}) above.
By substituting the expressions for ${\cal R}_{m,j}$ into Eq. (\ref{eq29}) and 
using the obtained solutions for the perturbative harmonic amplitude corrections 
(for various harmonics and orders) $n_{1,j}^{(1)}$, $u_{1,j}^{(1)}$, 
$\phi_{1,j}^{(1)} =\Psi_j$, $n_{2,j}^{(2)}$, $u_{2,j}^{(2)}$, $\phi_{2,j}^{(2)}$, 
$n_{2}^{(0)}$, $u_{2}^{(0)}$, and $\phi_{2}^{(0)}$, we end up after tedious and 
lengthy calculations with the equations
\begin{eqnarray}
\label{eq30}
   4 \omega_1 k_1 \frac{\partial^2 \Psi_1 }{\partial t_1 \partial x_1} 
  -\frac{k_1^2}{\omega_1^2} \frac{\partial^2 \Psi_1 }{\partial t_1^2}
  +(1-\omega_1^2) \frac{\partial^2 \Psi_1 }{\partial x_1^2}
  +2 i \frac{k_1^2}{\omega_1} \frac{\partial \Psi_1 }{\partial t_2} 
  +2 i k_1 (1-\omega_1^2) \frac{\partial \Psi_1 }{\partial x_2}
  \nonumber \\
  +\left( \tilde{Q}_{11} |\Psi_1|^2 +\tilde{Q}_{12} |\Psi_2|^2 \right) \Psi_1 =0,
  \\
 \label{eq31}
   4 \omega_2 k_2 \frac{\partial^2 \Psi_2 }{\partial t_1 \partial x_1} 
  -\frac{k_2^2}{\omega_2^2} \frac{\partial^2 \Psi_2 }{\partial t_1^2}
  +(1-\omega_2^2) \frac{\partial^2 \Psi_2 }{\partial x_1^2}
  +2 i \frac{k_2^2}{\omega_2} \frac{\partial \Psi_2 }{\partial t_2} 
  +2 i k_2 (1-\omega_2^2) \frac{\partial \Psi_2 }{\partial x_2}
  \nonumber \\
  +\left( \tilde{Q}_{21} |\Psi_1|^2 +\tilde{Q}_{22} |\Psi_2|^2 \right) \Psi_2 =0, 
\end{eqnarray}
where
\begin{equation}
\label{eq32}
   \tilde{Q}_{jj} =-2 \frac{k_j^3}{\omega_j} 
    \left( C_{u,2,j}^{(2)} +C_{u,2,j}^{(0)} \right)
   -k_j^2 \left( C_{n,2,j}^{(2)} +C_{n,2,j}^{(0)} \right)
   +2 c_2 \omega_j^2  \left( C_{\phi,2,j}^{(2)} +C_{\phi,2,j}^{(0)} \right) 
   +3 c_3 \omega_j^2,
\end{equation}
and
\begin{eqnarray}
\label{eq33}
   \tilde{Q}_{12} =-2 \frac{k_1^3}{\omega_1} C_{u,2,2}^{(0)}     
       -k_1  \frac{k_2}{\omega_2} \left( \omega_1 \frac{k_2}{\omega_2} +k_1 \right)
              \left( C_{u,2,p}^{(1)} +C_{u,2,m}^{(1)} \right)
               -k_1^2 C_{n,2,2}^{(0)} -\omega_1 k_1 \frac{k_2}{\omega_2} 
                 \left( C_{n,2,p}^{(1)} +C_{n,2,m}^{(1)} \right)
                         \nonumber \\
      +2 c_2 \omega_1^2  
      \left( C_{\phi,2,2}^{(0)} +C_{\phi,2,p}^{(1)} +C_{\phi,2,m}^{(1)} \right) 
      +6 c_3 \omega_1^2,
                        \\
\label{eq34}
   \tilde{Q}_{21} =-2 \frac{k_2^3}{\omega_2} C_{u,2,1}^{(0)}     
          -k_2  \frac{k_1}{\omega_1} \left( \omega_2 \frac{k_1}{\omega_1} +k_2 \right)
                 \left( C_{u,2,p}^{(1)} +C_{u,2,m}^{(1)} \right)
          -k_2^2 C_{n,2,1}^{(0)} -\omega_2 k_2 \frac{k_1}{\omega_1} 
                  \left( C_{n,2,p}^{(1)} +C_{n,2,m}^{(1)} \right)
                         \nonumber \\
       +2 c_2 \omega_2^2  
        \left( C_{\phi,2,1}^{(0)} +C_{\phi,2,p}^{(1)} +C_{\phi,2,m}^{(1)} \right) 
      +6 c_3 \omega_2^2.  
\end{eqnarray}
Note that the terms proportional to the variables $n_{2,j}^{(1)}$, 
$u_{2,j}^{(1)}$, and $\phi_{2,j}^{(1)}$ were eliminated with the help 
of Eqs. (\ref{eq12.3}).

Applying the second-order compatibility conditions to simplify Eqs. (\ref{eq30}) 
and (\ref{eq31}), we obtain a pair of coupled NLS equations in the form of Eqs. (\ref{eq35})-(\ref{eq36}).

\section{Coefficients in the calculation -- analytical expressions}

\begin{eqnarray}
   C_{n,2,j}^{(2)} =\frac{(k_j^2 +c_1)}{6 k_j^2} 
       \left[3 (4 k_j^2 +c_1) (k_j^2 +c_1) -2 c_2 \right],
\\
\label{eq131}
   C_{u,2,j}^{(2)} =\frac{\sqrt{k_j^2 +c_1}}{6 k_j^2} 
       \left[3 (k_j^2 +c_1) (2 k_j^2 +c_1) -2 c_2 \right],
\\
   C_{\phi,2,j}^{(2)} =\frac{1}{6 k_j^2} 
       \left[3 (k_j^2 +c_1)^2 -2 c_2 \right],
\end{eqnarray}

\begin{eqnarray}
   C_{n,2,\pm}^{(1)} =\frac{1}{D_\pm} \left\{ \left[ (k_1 \pm k_2)^2 +c_1 \right] 
        \left[ (k_1 \pm k_2) f_2^{(\pm)} +(\omega_1 \pm \omega_2) f_1^{(\pm)} \right] 
        +(k_1 \pm k_2)^2 f_3^{(\pm)}
   \right\},
   \nonumber \\
   \label{eq132}
   C_{u,2,\pm}^{(1)} =\frac{1}{D_\pm} \left\{ (k_1 \pm k_2) 
       \left[ f_1^{(\pm)} +(\omega_1 \pm \omega_2) f_3^{(\pm)} \right]
         +(\omega_1 \pm \omega_2) \left[ (k_1 \pm k_2)^2 +c_1 \right] f_2^{(\pm)}
   \right\},
   \\
   C_{\phi,2,\pm}^{(1)} =\frac{1}{D_\pm} \left\{ 
   (\omega_1 \pm \omega_2) \left[ f_1^{(\pm)} +(\omega_1 \pm \omega_2) f_3^{(\pm)} \right] 
   +(k_1 \pm k_2) f_2^{(\pm)} 
   \right\},
   \nonumber
\end{eqnarray}
where 
\begin{equation}
\label{eq133}
   D_\pm =(k_1 \pm k_2)^2 -(\omega_1 \pm \omega_2)^2 \left[ (k_1 \pm k_2)^2 +c_1 \right].
\end{equation}
The quantities $f_m^{(\pm)}$ ($m=1,2,3$) are defined in Eqs. (\ref{eq12.6})-(\ref{eq12.11}).

\begin{equation}
\label{eq134}
   C_{n,2,j}^{(0)} =\left( c_1 C_{\phi,2,j}^{(0)} +2 c_2 \right), \qquad
   C_{u,2,j}^{(0)} =\frac{1}{v_{g,j}} 
       \left[ C_{\phi,2,j}^{(0)} +\frac{k_j^2}{\omega_j^2} \right], \qquad
   C_{\phi,2,j}^{(0)} =\rho_j \frac{1}{1 -c_1 v_{g,j}^2},
\end{equation}
where 
\begin{equation}
\label{eq135}
   \rho_j =-( k_j^2 +c_1 ) -2 c_1 +2 c_2 v_{g,j}^2. 
\end{equation}

\end{document}